\documentclass[prd,showpacs,showkeys,amsmath,amssymb,superscriptaddress,preprintnumbers,nofootinbib]{revtex4}
\usepackage{graphicx}% Include figure files
\usepackage{bm}% bold math

\normalbaselines
\newcommand{\arccosh}{\mathop{\rm arccosh}\nolimits}

\newcommand{\Ref}[1]{(\ref{#1})}
\newcommand{\cqg}{Classical Quantum Gravity\ }
\newcommand{\grg}{Gen. Relativ. Gravit.\ }
\newcommand{\jmp}{J. Math. Phys. (N.Y.)\ }

\begin{document}

\title{Nonminimal coupling for the gravitational and electromagnetic
fields:
\\ Traversable electric wormholes}
\author{Alexander B. Balakin}
\email{Alexander.Balakin@ksu.ru} \affiliation{Department of
General Relativity and Gravitation, Kazan State University,
Kremlevskaya str. 18, Kazan 420008, Russia.}
\author{Jos\'e P. S. Lemos}%
   \email{joselemos@ist.utl.pt}
   \affiliation{Centro Multidisciplinar de Astrof\'{\i}sica-CENTRA
Departamento de F\'{\i}sica,
Instituto Superior T\'ecnico-IST,\\
Universidade T\'{e}cnica de Lisboa-UTL, Avenida Rovisco Pais 1,
1049-001 Lisboa, Portugal }
\author{Alexei E. Zayats}
\email{Alexei.Zayats@ksu.ru} \affiliation{Department of General
Relativity and Gravitation, Kazan State University, Kremlevskaya
str. 18, Kazan 420008, Russia.}
%
%\preprint{\bf version 1.3}
%             %  but any date may be explicitly specified
%
\begin{abstract}
We discuss new exact solutions of a three-parameter nonminimal
Einstein-Maxwell model. The solutions describe static spherically
symmetric objects with and without center, supported by an
electric field nonminimally coupled to gravity. We focus on a
unique one-parameter model, which admits an exact solution for a
traversable electrically charged wormhole connecting two
universes, one asymptotically flat the other asymptotically de
Sitter ones.  The relation between the asymptotic mass and charge
of the wormhole and its throat radius is analyzed.  The wormhole
solution found is thus a nonminimal realization of Wheeler's idea
about charge without charge and shows that, if the world is
somehow nonminimal in the coupling of gravity to electromagnetism,
then wormhole appearance, or perhaps construction, is possible.
\end{abstract}
%                             % Classification Scheme.
%Use showkeys class option if keyword
%                              %display desired
\maketitle

\section{Introduction}

The wormhole concept was invented by Wheeler (see, e.g.,
\cite{Wheeler}) to provide a mechanism for having charge without
charge, since in a such a spacetime without a center, the field
lines seen in one part of the Universe could thread the throat and
reappear in other part. The idea, was further extended by Morris
and Thorne to allow, not only field lines, but also observers to
travel through the throat \cite{MorTho}. By having invoked an
arbitrarily technologically advanced civilization, this work
\cite{MorTho} initiated the engineering of wormhole construction,
theoretically, and the study of wormholes as topological bridges
joining two different spacetimes has since then attracted
extraordinary attention in this modern context (see, e.g.,
\cite{VisserBook} and references therein).  The main feature of
traversable wormhole physics is the fact that the matter threading
the wormhole throat should possess exotic properties, one of them
being the violation of the null energy condition \cite{HocVis}. To
provide the existence of wormholes one should either include some
hypothetical forms of matter into the model, or introduce
interactions of a new type. Various models, admitting the required
violation of the null energy condition, have been considered in
the literature, among them thin shells in a cosmological constant
background \cite{LambdaWH1,LambdaWH2}, scalar fields
\cite{scalarfields}, wormhole solutions in semiclassical gravity
\cite{semiclas}, solutions in Brans-Dicke theory \cite{Nan-etal},
wormholes on the brane \cite{wormholeonbrane}, wormholes supported
by matter with an exotic equation of state, namely, phantom energy
\cite{phantom}, the generalized Chaplygin gas \cite{chaplygin},
tachyon matter \cite{tachyon}, nonlinear electrodynamics
\cite{AreLob}, and other cases.

Now, when one considers nonminimal coupling of gravity with
vector-type fields, such as Maxwell, Yang-Mills or Proca fields,
new interesting possibilities appear. Nonminimal phenomena, i.e.,
phenomena induced by the interaction between curvature, or
gravity, and other fields can be characterized on the one hand by
unusual effective energy conditions, and on the other hand, allow
one to exclude exotic substrates. In \cite{BaSuZa} an exact
solution of the nonminimal Einstein--Yang-Mills model was obtained
which demonstrates that an $SU(2)$ symmetric gauge field
nonminimally coupled to curvature can support a traversable
wormhole. This is a nonminimal Wu-Yang magnetic wormhole, and the
throat joins two asymptotically flat regions. The corresponding
spacetime has no center and this model could be an illustration of
Wheeler's idea about ``charge without charge'' \cite{Wheeler}.
Following this idea we now intend to consider an electrically
charged object in the context of a spacetime without a center. A
few wormhole models are known in which the electric charge is
spread on the spherical shell \cite{thinshell}, or the throat is
filled with some nonminimal and ghostlike scalar field
\cite{charge}. Here, we find an exact solution of a nonminimal
traversable wormhole, which contains neither electric charge on
spherical shells, either thin or thick, nor scalar fields,
nevertheless, possesses a static spherically symmetric electric
field which is charged from the point of view of a distant
observer.  Thus, our goal is twofold.  We present explicitly a
nonminimal realization of Wheeler's idea about charge without
charge, and we show that, if the world is somehow nonminimal in
the coupling of gravity to electromagnetism, then wormhole
appearance, or perhaps construction by an absurdly advanced
civilization \cite{LambdaWH1}, is possible.

For this purpose we address a nonminimal Einstein-Maxwell theory,
which has been earlier elaborated in detail in both linear (see,
e.g., \cite{FaraR,HehlObukhov}) and nonlinear (see \cite{BL05})
versions. The model linear in the spacetime curvature and
quadratic in the Maxwell tensor \cite{BL05} contains three
nonminimal coupling constants $q_1$, $q_2$, and $q_3$. These
quantities have the dimensionality of area and characterize the
cross terms in the Lagrangian linking the Maxwell field $F_{ik}$
and terms linear in the Ricci scalar $R$, Ricci tensor $R_{ik}$,
and Riemann tensor $R_{ikmn}$, respectively. These parameters are
a priori free ones but can acquire specific values in certain
effective field theories.  The first example of a calculation of
the three coupling parameters was based on one-loop corrections to
quantum electrodynamics in curved spacetime, a direct and
nonphenomenological approach considered by Drummond and Hathrell
\cite{Drum}.  In another instance, Buchdahl \cite{Buchdahl} and
then M\"uller-Hoissen \cite{MH} obtained a nonminimal
Einstein-Maxwell model from dimensional reduction of the
Gauss-Bonnet action. This model contains one coupling parameter.

Based on these works, and specially on \cite{BL05}, solutions,
which described nonminimal electrically charged objects with and
without centers, characterized by two sets of relations for
nonminimal coupling parameters, namely, $q_1+q_2+q_3=0$,
$2q_1+q_2=0$, and $q_1+q_2=0$, $q_3=0$, were obtained
\cite{BBL08}.  However, none of these solutions can be used to
construct traversable wormholes. In this paper we formulate a new
nonminimal model, for which the coupling parameters satisfy the
relations, $3q_1+q_2=0$, $q_3=0$. Exact solutions of this model
are shown to admit the existence of nonminimal traversable
electrically charged wormholes.  Comparing these new results with
the ones obtained in \cite{BaSuZa} we would like to emphasize the
following features. First, here we deal with an electric field
instead of a magnetic one; second, in \cite{BaSuZa} we used the
basic conditions $12q_1+4q_2+q_3=0$, $q_3\neq0$; third, the
spacetime is now nonsymmetric, i.e., the throat joins one
asymptotically flat region to another asymptotically de Sitter
region. In \cite{BronMD} the authors have suggested the name
``black universes'' to nonsymmetric wormhole spacetime
configurations for which the first asymptotics is flat and the
second one is of cosmological type.

The paper is organized as follows: In Section \ref{main}, we quote
the details of the three-parameter nonminimal Einstein-Maxwell
model and formulate the corresponding key equations for the
electric and gravitational fields. Furthermore, we introduce a new
one-parameter nonminimal model, given by the conditions
$3q_1+q_2=0$, $q_3=0$, derive a key (cubic) equation for the
electric field $E$ and obtain the metric functions $\sigma$, $N$
in terms of $E$.  In Section III, we discuss three solutions of
the equation for the electric field. We focus on the
Coulombian-type solution: This is the solution to which the others
should be compared.  In Section IV, we obtain exact solutions with
a center. In Section V, we thoroughly discuss wormhole solutions.
Section VI contains the conclusions.

\section{Nonminimally extended Einstein-Maxwell theory}\label{main}

\subsection{Master equations}

Details of three-parameter nonminimal Einstein-Maxwell theory can
be found in the papers \cite{BL05,BBL08}; here we extract the main
elements of this model only. The action functional is of the form
\begin{equation}
S_{{\rm NMEM}} = \int d^4 x \sqrt{-g}\
\left[\frac{R}{\kappa}+\frac{1}{2}F_{ik} F^{ik}+\frac{1}{2}\,
{\chi}^{ikmn}F_{ik} {F}_{mn}\right]\,.\label{act}%
\end{equation}%
Here, $g = {\rm det}(g_{ik})$ is the determinant of the metric
tensor $g_{ik}$, $R$ is the Ricci scalar, $\kappa$ is the
gravitational constant. The Latin indices run from 0 to 3. The
Maxwell tensor $F_{ik}$ is expressed, as usual, in terms of a
potential four-vector $A_k$
\begin{equation}
F_{ik} = \nabla_i A_k - \nabla_k A_i \,, \label{F}%
\end{equation}
where the symbol $\nabla_i$ denotes the covariant derivative. The
tensor ${\chi}^{ikmn}$ is defined as follows (see \cite{BL05}):
\begin{equation}
{\chi}^{ikmn} \equiv \frac{q_1}{2}R\,(g^{im}g^{kn}-g^{in}g^{km}) +
\frac{q_2}{2}(R^{im}g^{kn} - R^{in}g^{km} + R^{kn}g^{im}
-R^{km}g^{in}) + q_3 R^{ikmn} \,, \label{sus}
\end{equation}%
where $R^{ik}$ and $R^{ikmn}$ are the Ricci and Riemann tensors,
respectively, and $q_1$, $q_2$, $q_3$ are the phenomenological
parameters describing the nonminimal coupling of electromagnetic
and gravitational fields. The variation of the action functional
with respect to potential $A_i$ yields
\begin{equation}
\nabla_k H^{ik}=0\,, \quad H^{ik}\equiv F^{ik} +
{\chi}^{ikmn}F_{mn} \,, \label{HikR}
\end{equation}
where $H^{ik}$ is the nonminimal excitation tensor \cite{Maugin}.
In a similar manner, the variation of the action functional with
respect to the metric yields
\begin{equation}
R_{ik} - \frac{1}{2} R \ g_{ik} = \kappa\,T^{({\rm eff})}_{ik} \,.
\label{Ein}
\end{equation}
The effective stress-energy tensor $T^{({\rm eff})}_{ik}$ can be
divided into four parts:
\begin{equation}
T^{({\rm eff})}_{ik} =  T^{(M)}_{ik} + q_1 T^{(I)}_{ik} + q_2
T^{(II)}_{ik} + q_3 T^{(III)}_{ik} \,. \label{Tdecomp}
\end{equation}
The first term $T^{(M)}_{ik}$:
\begin{equation}
T^{(M)}_{ik} \equiv \frac{1}{4} g_{ik} F_{mn}F^{mn} -
F_{in}F_{k}^{\ n} \,, \label{TYM}
\end{equation}
is a stress-energy tensor of the pure electromagnetic field. The
definitions of other three tensors are related to the
corresponding coupling constants $q_1$, $q_2$, $q_3$:
\begin{equation}%
T^{(I)}_{ik} = R\,T^{(M)}_{ik} -  \frac{1}{2} R_{ik} F_{mn}F^{mn}
+ \frac{1}{2} \left[ \nabla_{i} \nabla_{k} - g_{ik} \nabla^l
\nabla_l \right] \left[F_{mn}F^{mn} \right] \,, \label{TI}
\end{equation}%

\[%
T^{(II)}_{ik} = -\frac{1}{2}g_{ik}\biggl[\nabla_{m}
\nabla_{l}\left(F^{mn}F^{l}_{\ n}\right)-R_{lm}F^{mn} F^{l}_{\ n}
\biggr] - F^{ln}
\left(R_{il}F_{kn} + R_{kl}F_{in}\right)-R^{mn}F_{im} F_{kn} {} \]%
\begin{equation}%
{} - \frac{1}{2} \nabla^m \nabla_m \left(F_{in} F_{k}^{ \
n}\right)+\frac{1}{2}\nabla_l \left[ \nabla_i \left( F_{kn}F^{ln}
\right) + \nabla_k \left(F_{in}F^{ln} \right) \right] \,,
\label{TII}
\end{equation}%

\begin{equation}%
T^{(III)}_{ik} = \frac{1}{4}g_{ik} R^{mnls}F_{mn}F_{ls}-
\frac{3}{4} F^{ls} \left(F_{i}^{\ n} R_{knls} + F_{k}^{\
n}R_{inls}\right) - \frac{1}{2}\nabla_{m} \nabla_{n} \left[
F_{i}^{ \ n}F_{k}^{ \ m} + F_{k}^{ \ n} F_{i}^{ \ m} \right] \,.
\label{TIII}
\end{equation}%
One may check directly that the tensor $T^{({\rm eff})}_{ik}$
satisfies the equation $\nabla^k T^{({\rm eff})}_{ik} =0$.

\subsection{Spherically symmetric model with electric charge}
Below we consider the nonminimally extended Einstein-Maxwell
equations, given in Eqs.~(4)-(10),  for the case of a static
spherically symmetric spacetime, given by the metric
\begin{equation}\label{metrica}
ds^2=\sigma^2Ndt^2-\frac{dr^2}{N}-Y^2 \left( d\theta^2 +
\sin^2\theta d\varphi^2 \right) \,,
\end{equation}
where $N$, $\sigma$, and $Y$ are functions of the radial variable
$r$ only satisfying the asymptotic flatness conditions when $r\to
+\infty$
\begin{equation}
\sigma(r)\to 1\,,\quad N(r)\to 1\,,\quad Y(r)\sim r\,.
\label{asympCond}
\end{equation}
In this paper, asymptotic flatness means that we do not consider
the case $Y'(r)=0$ identically. Let us assume also that the
electromagnetic field inherits the static and spherical
symmetries. Then the potential four-vector of the electric field
$A_i$ and the Maxwell tensor have the form
\begin{equation}
\label{A} A_i=A_0(r)\delta^{0}_i\,, \quad F_{ik}=A_0'(r)
\left(\delta_i^{\,r}\delta_k^0-\delta_k^0\delta_i^{\,r}\right)\,,
\end{equation}
where a prime denotes the derivative with respect to $r$. To
characterize the electric field, it is useful to introduce a new
scalar quantity $E(r)$ as
\begin{equation}
E^2(r)=-\frac12 F_{ik}F^{ik} \label{AB}
\end{equation}
Then the electric field squared is $\left(A_0'\right)^2$ from
which one obtains in turn $F_{r0}=-\sigma(r)E(r)$.

The Maxwell equations \Ref{HikR} give only one nontrivial
equation, namely,
\begin{equation}\label{Meq}
      \left[Y^2E(1+2{{\chi}^{r0}}_{r0})\right]'=0\,,
\end{equation}
which can be integrated immediately to give
\begin{equation}\label{Emain}
    E = \frac{Q}{Y^2(1+2{{\chi}^{r0}}_{r0})}\,,
\end{equation}
where $Q$ is a constant to be associated with an electric charge
of the object. Supposing the spacetime to be asymptotically flat,
i.e., $R_{iklm}(r\to +\infty)= 0$, one can see that \Ref{Emain}
yields asymptotically the Coulomb law $E(r)\to Q/r^2$. The
nonminimal excitation tensor $H^{ik}$ given in (\ref{HikR}) with
$\chi^{ikmn}$ given by (\ref{sus}) has only one nonvanishing
component $H^{r0}$. Let us introduce an electric excitation scalar
as follows
\begin{equation}\label{D}
      D\equiv
      \sqrt{-\frac{1}{2}H_{ik}H^{ik}}=\sqrt{-H_{r0}H^{r0}}=
E(1+2{{\chi}^{r0}}_{r0})\,.
\end{equation}
According to (\ref{Emain}) the electric excitation takes a simple
form
\begin{equation}
      D=\frac{Q}{Y^2}\,,
\end{equation}
i.e., $D$ also tends to zero at $r\to +\infty$.

\subsection{Key equations}

\subsubsection{Preliminaries}
Analogously to the static spherically symmetric model in minimal
electrodynamics there are two independent Einstein equations only,
e.g.,
\begin{equation}\label{19}
G_0^{\,0}-G_r^{\,r}=\kappa \left(T_0^{\,0({\rm
eff})}-T_r^{\,r({\rm eff})}\right)\,, \quad G_0^{\,0}=\kappa
T_0^{\,0({\rm eff})}\,.
\end{equation}
The first one gives the equation linking the function $\sigma(r)$
with $E(r)$ and $Y(r)$,
\begin{equation}\label{sgen}
      \frac{\sigma'}{\sigma}=\frac{\kappa
(q_1+q_2+q_3)\left[Y^2(EE')'+YY''E^2\right]+2\kappa(q_2+2q_3)YY'EE'+
\kappa q_3Y'^2E^2-YY''}
      {Y\left[\kappa (q_1+q_2+q_3)\left(YEE'+Y'E^2\right)-Y'\right]}\,.
\end{equation}
It is convenient to rewrite the second equation in (\ref{19})
using \Ref{Emain} as
\begin{equation}
      G_0^{\,0}-\kappa T_0^{\,0({\rm eff})}+2\kappa E^2{{\chi}^{r0}}_{r0}=
\kappa E^2\left(\frac{Q}{Y^2E}-1\right).
\end{equation}
For generic coupling constants this equation yields
\begin{eqnarray}
      N'Y\left[\kappa
      (q_1+q_2+q_3)(YEE'+Y'E^2)-Y'\right]+N\left\{2\kappa (q_1+q_2+q_3)
      \left[Y^2(EE')'+YY''E^2\right]+{}\right.\nonumber\\ \left.{}+
2\kappa EE'YY'(2(q_1+q_2+q_3)+q_2+2q_3)+\kappa
      ((q_1+q_2+q_3)+q_3)E^2Y'^2-Y'^2-2YY''\right\}+{}\nonumber\\{}+1+
      \frac{\kappa
      E^2Y^2}{2}-\kappa q_1 E^2-\kappa Q E=0\,.\label{EN0}
\end{eqnarray}
The three Eqs.~\Ref{Emain}, \Ref{sgen} and \Ref{EN0} form the key
system of equations determining the four unknown functions
$\sigma$, $N$, $E$ and $Y$. One of these functions is arbitrary.
Generally, one has $\sigma'(r)$ and $N'(r)$ given in
Eqs.~\Ref{sgen} and \Ref{EN0}, and put both into \Ref{Emain}. This
procedure yields
\begin{gather}
     2 N E
\left[\kappa^2(q_1+q_2+q_3)^2(3q_1+q_2)YE^2(EE''YY'-EE'YY''-YY'E'^2+EE'Y'^2)
-{}\right.\nonumber\\
      {}-\kappa (q_1+q_2+q_3)(3q_1+q_2-q_3)Y^2(EE''Y'-EE'Y''+Y'E'^2)-{}
\nonumber\\
\left.{}-\kappa (q_1+q_2+q_3)(3q_1+q_2-4q_3)YY'^2EE'+\kappa
      q_3(q_1+q_2+q_3)Y'^3E^2-q_3Y'^3\right] = {} \nonumber\\
{} =
\left\{\kappa^2E^4E'(q_1+q_2+q_3)[(2q_1+q_2)Y^3-2q_1Y(2q_1+q_2+(q_1+q_2+q_3))]
\right.\nonumber\\
-\kappa^2YQE^3E'(q_1+q_2+q_3)(4q_1+2q_2+(q_1+q_2+q_3))-{}\nonumber\\
      {}-\kappa E^2E'(q_1+q_2+q_3)(Y^3-2Y(3q_1+q_2))+\kappa Y QEE'
(q_1+q_2+q_3)-\kappa^2 Y'E^5(q_1+q_2+q_3) q_1
(Y^2+2(q_2+q_3))+{}\nonumber\\
      \left.{}+\kappa^2QY'E^4(q_1+q_2+q_3)(q_1-q_2-q_3)
-\kappa Y'E^3(q_2+2q_3)(Y^2-2q_1)+2\kappa
QY'E^2(q_2+2q_3)+{}\right.
\nonumber\\
      \left.{}+Y'E(Y^2-2q_3)-QY'\right\} \,, \label{NYgen}
\end{gather}
thus giving $N$ as a function of $E$, $E'$, $E''$, $Y$, $Y'$ and
$Y''$. In principle, putting $N$ from Eq.~(\ref{NYgen}) in
Eq.~\Ref{EN0} one can obtain a key equation linking $E(r)$ and
$Y(r)$. This equation happens to be nonlinear and contains third
order derivatives $E'''$ and $Y'''$. Although there are no
explicit exact solutions for arbitrary $q_1$, $q_2$, $q_3$, one
can estimate the behavior of the integral curves using a
decomposition with respect to $Y^{-1}$. For instance, when the
model is asymptotically flat, the decompositions start as
\begin{gather}
E=\frac{Q}{Y^2}\,\left(1+\frac{4q_3M}{Y^3}-
\frac{\kappa(q_2+3q_3)Q^2}{Y^4}\right)+\dots\,,\\
\sigma = Y'\left[1+\frac{\kappa(10q_1+6q_2+3q_3)Q^2}{4Y^4}+
\dots \right]\,,\label{sigmaasymp}\\
N = \frac{1}{Y'^2} \left[1-\frac{2M}{Y}+\frac{\kappa
Q^2}{2Y^2}-\frac{\kappa(4q_1+3q_2+2q_3)Q^2}{Y^4}+\dots
\right]\,,\label{Nasymp}
\end{gather}
where the constant $M$ is an asymptotic mass of the object.

There are two special cases arising when the left-hand side of
Eq.~(\ref{NYgen}) vanishes so that the function $N$ cannot be
found from this equation directly. The first one is the model with
$q_1+q_2+q_3= 0$ and $q_3=0$, which has been studied in
\cite{BBL08}, the second relates to the model with $3q_1+q_2=0$
and $q_3=0$. Since the model with $q_1+q_2+q_3=0$ and $q_3=0$ has
been studied in \cite{BBL08} we briefly mention that this
one-parameter model can be characterized by the relationships
$q_1=-q$, $q_2=q$, $q_3=0$, for some $q$, and (\ref{NYgen}) is
satisfied, when the right-hand side of this equation also
vanishes, i.e.,
\begin{equation}\label{0E1}
\kappa q E^3 (Y^2+2q)-2\kappa q Q E^2 -E Y^2+Q = 0 \,.
\end{equation}
The function $\sigma$ can be readily found as
\begin{equation}\label{s1}
    \sigma = Y'\exp(-\kappa q E^2)\,. \\
\end{equation}
When the function $Y$ coincides with the radius $r$, i.e.,
$Y(r)=r$, we obtain indeed the model which has been investigated
in the paper \cite{BBL08}.  On the other hand the model with
$3q_1+q_2=0$ and $q_3=0$ is also very interesting and will be
studied in detail here.

\subsubsection{One-parameter model with $3q_1+q_2=0$ and
$q_3=0$}\label{C2}

The one-parameter model with $3q_1+q_2=0$ and $q_3=0$ is
characterized by the relationships $q_1=-q$, $q_2=3q$, $q_3=0$,
for some $q$. The susceptibility tensor is now proportional to the
difference of the Riemann ($R_{ikmn}$) and Weyl ($C_{ikmn}$)
tensors
\begin{equation}
      {\chi}_{ikmn}=3q\left(R_{ikmn}-C_{ikmn}\right)\,.
\end{equation}
Equation \Ref{NYgen} is satisfied when
\begin{equation}\label{0E2}
-\left[Y'-2\kappa q(YEE'+Y'E^2)\right][\kappa q E^3
(Y^2+6q)-4\kappa q Q E^2 -E Y^2+Q]= 0 \,.
\end{equation}
Again, one has two variants to obtain exact solutions. The first
one can be realized, when the first bracket vanishes, and we
obtain
\begin{equation}\label{0E3}
Y'-2\kappa q(YEE'+Y'E^2) = 0  \ \ \rightarrow  \ \ Y^2 = \frac{\rm
const}{1-2\kappa q E^2}\,.
\end{equation}
The minimal limit $q \to 0$ relates to the model $Y \to {\rm
const}$ instead of $Y \to r$,  and we do not consider here such a
model. When the second bracket in (\ref{0E2}) vanishes, we obtain
the following cubic equation for the electric field
\begin{equation}%\label{00E0}
    \kappa q E^3 (Y^2+6q)-4\kappa q Q E^2 -E Y^2+Q=0\,. \label{00E0}
\end{equation}
This equation differs from (\ref{0E1}) by the numerical
multipliers in front of $q$. Surprisingly, the equation for the
function $\sigma(r)$ [see Eq.~(\ref{sgen})] can be now explicitly
resolved in terms of $E$ and $Y$ as
\begin{equation}\label{s0}
    \sigma = Y'-2\kappa q \left(E^2Y'+YEE'\right)\,. \\
\end{equation}
In order to find the function $N(r)$ we consider Eq.~(\ref{EN0}),
which gives
\begin{equation}
    N = \frac{1}{\sigma^2Y}\left\{\int dr\, \sigma
    \left(1+\frac{\kappa E^2 Y^2}{2}+\kappa q E^2-
\kappa Q E\right)+{\rm const}\right\}\,. \label{N88}
\end{equation}
Thus, any solution $E(Y)$ of the cubic equation \ref{00E0}) for
the electric field gives us a new exact solution of this
nonminimal model, since (\ref{s0}) gives immediately the function
$\sigma(r)$ and (\ref{N88}) gives the function $N(r)$ in
quadratures.

\subsubsection{Dimensionless quantities and equations}

In order to analyze  Eqs.~(\ref{00E0})-(\ref{N88}), and in
particular the cubic equation (\ref{00E0}) for the electric field,
let us introduce the following dimensionless quantities (see
\cite{BBL08})
\begin{equation}\label{dimless}
r_Q=\sqrt{\kappa Q^2/2} \,, \quad  E_Q=Q/r^2_Q \,, \quad a =
\frac{2q}{r_Q^2} \,, \quad  \rho=\frac{r}{r_Q}\,, \quad y =
\frac{Y}{r_Q} \,, \quad  Z=\frac{E}{E_Q}\,.
\end{equation}
In these terms the Coulombian branch of Eq.~(\ref{00E0})
corresponds to the solution with the asymptotic behavior $Z \sim
\rho^{-2}$ at $\rho\to +\infty$. Then the key equations
\Ref{00E0}--\Ref{N88} take the form
\begin{equation}\label{E}
    a\,Z^3 (y^2+3a)-4a\,Z^2 -Z y^2+1=0\,,
\end{equation}
\begin{equation}\label{s}
    \sigma = \frac{dy}{d\rho}-2a \left(Z^2\frac{dy}{d\rho}+
yZ\frac{dZ}{d\rho}\right)\,,
\end{equation}
\begin{equation}\label{N}
    N = \frac{1}{\sigma^2y}\left\{\int d\rho\, \sigma
    \left(1+Z^2 y^2+a\,Z^2-2Z\right)+{\rm const}\right\}\,.
\end{equation}
The dimensionless parameter $a$ is a guiding parameter for the key
equation (\ref{E}). When $a=0$, then $q=0$ and the model is
minimal.

\subsubsection{Solution with $a<0$}

\begin{figure}[h]
\halign{\hfil#\hfil&\qquad\hfil#\hfil&\qquad\hfil#\hfil\cr
\includegraphics[height=3.8cm]{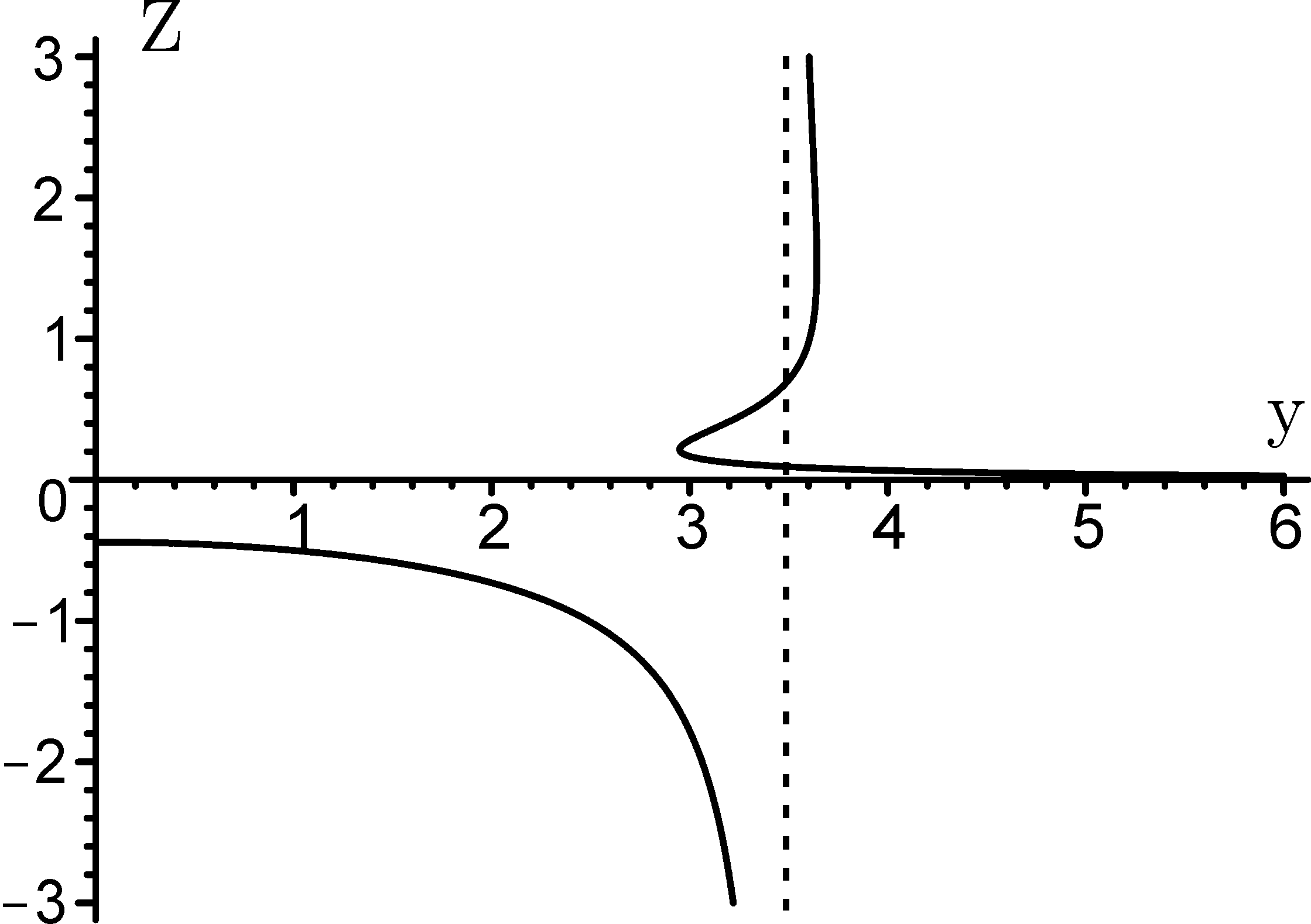}&
\includegraphics[height=3.8cm]{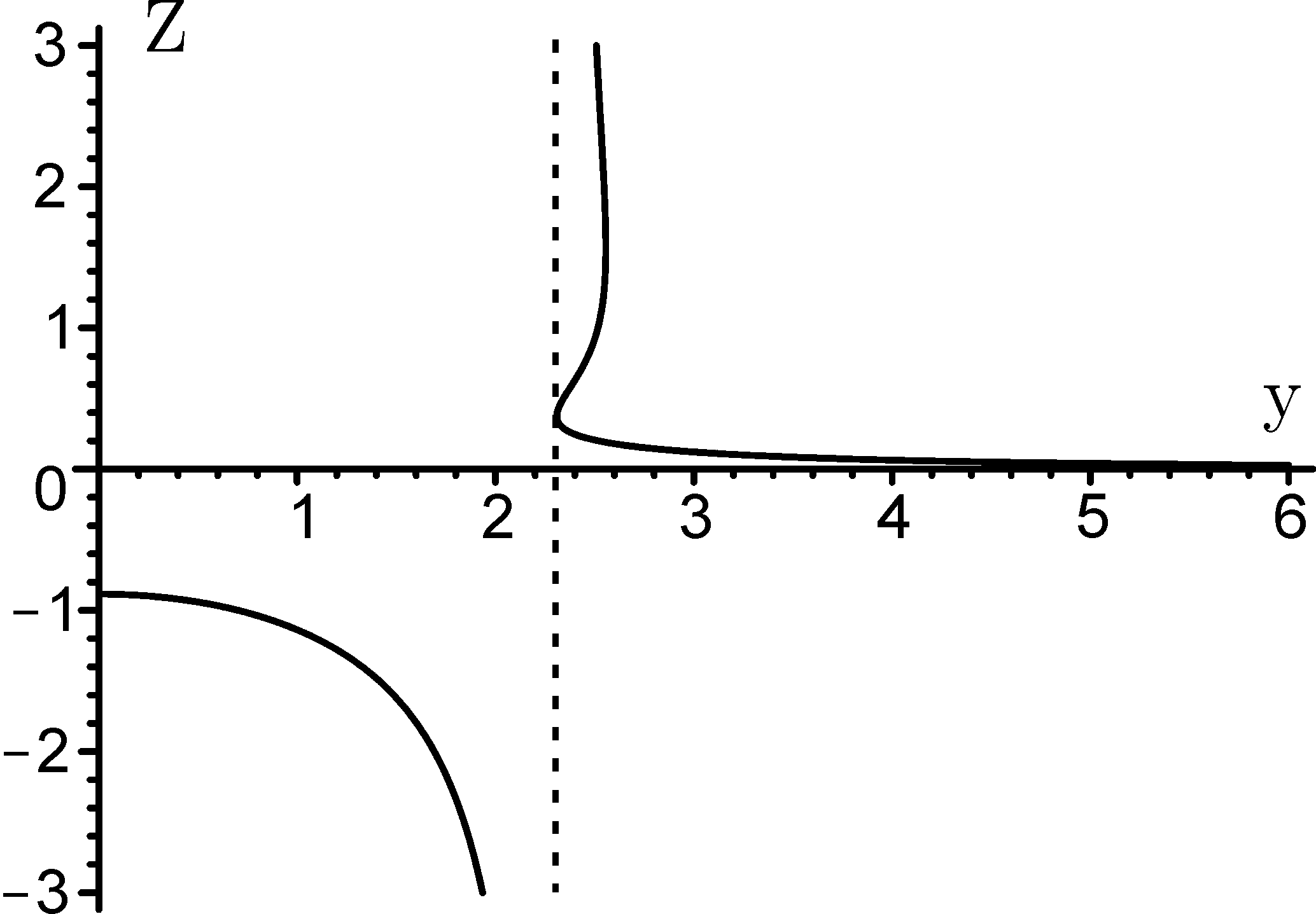}&
\includegraphics[height=3.8cm]{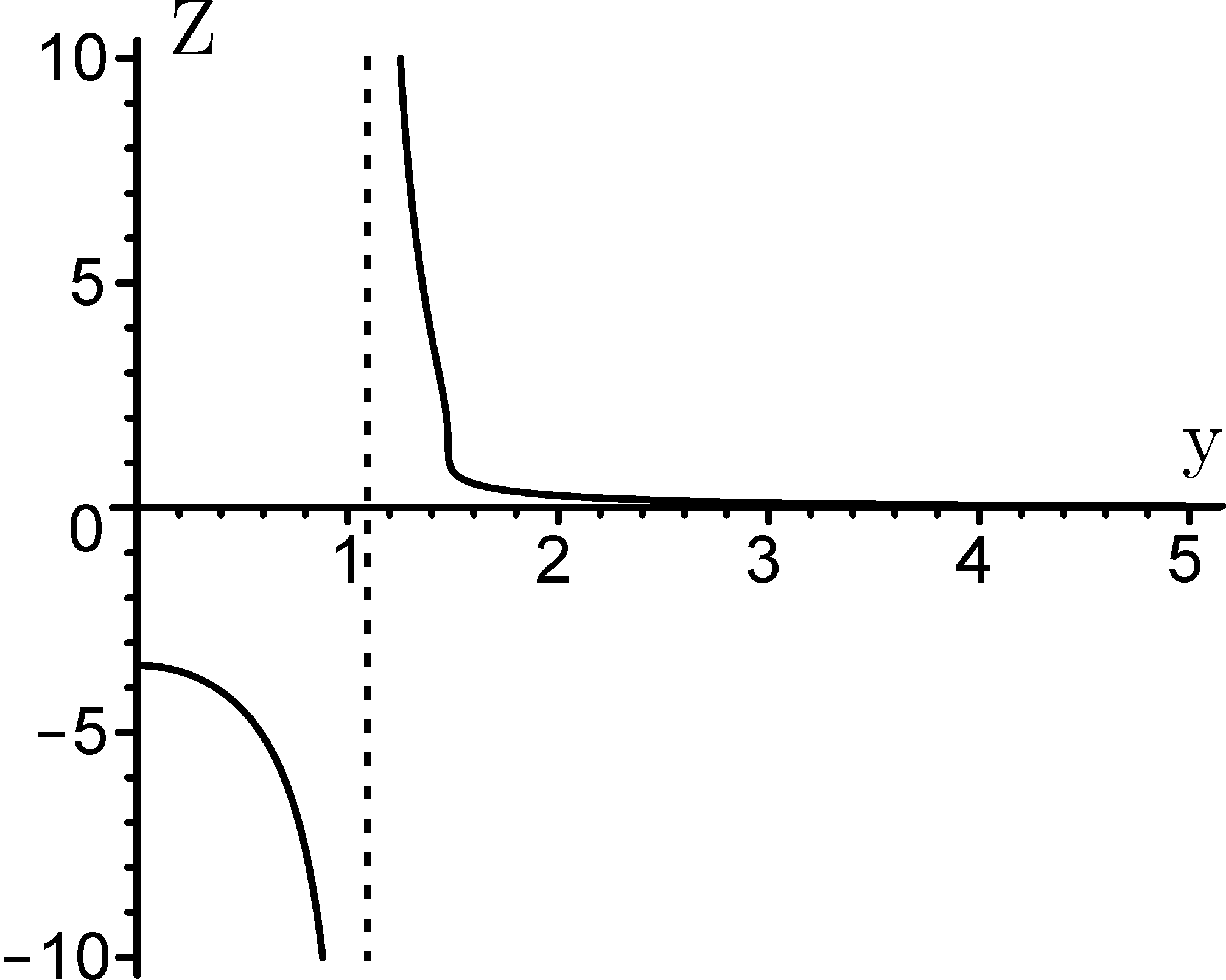}\cr
\small a) $a<-16/9$ & \small b) $a=-16/9$ & \small c) $-16/9<a<0$
\cr} \caption{Plots of the rescaled electric field $Z(y)=E/E_Q$ as
function of $y=Y/r_Q$ for $a<0$. All curves have a vertical
asymptote at $y=\sqrt{-3a}$. (a) When $-\infty<a<-16/9$, the curve
crosses the asymptote twice. (b) When $a=-16/9$, the curve touches
the asymptote at one point. (c) When $-16/9<a<0$, the curve does
not cross the asymptote. For all cases the function $Z(y)$ is
infinite, the singularity point being situated at $y=0$, when
$a=0$ only. When $a\to -0$, the vertical asymptote coincides with
the coordinate line $y=0$ and the right branch of the curve (c)
converts into the standard Coulombian curve
$Z(y)=1/y^2$.}\label{nega}
\end{figure}

Before we enter into the really interesting solutions, we mention
the solutions with $a<0$.  The expression $y^2 + 3a$, the
coefficient of $Z^3$ in the Eq.~(\ref{E}), can vanish, when $a<0$;
thus, the corresponding solution $Z(y)$ is singular. The equation
$y=\sqrt{-3a}$ describes a vertical asymptote. There exists one
critical value of the guiding parameter $a$, namely, $a=-16/9$,
for which the point characterizing by the condition
$\frac{dZ}{dy}=\infty$, belongs to this vertical asymptote. The
typical behavior of electric field at $a<0$ is presented on
Fig.~\ref{nega}.  Anyway, we will consider regular solutions only,
thus assuming that $a>0$ from now on.

\section{Electric field: The
Coulombian and non-Coulombian solutions} \label{The Coulombian
solution}

\begin{figure}[h]
\halign{\hfil#\hfil&\qquad\hfil#\hfil&\qquad\hfil#\hfil\cr
\includegraphics[height=4.0cm]{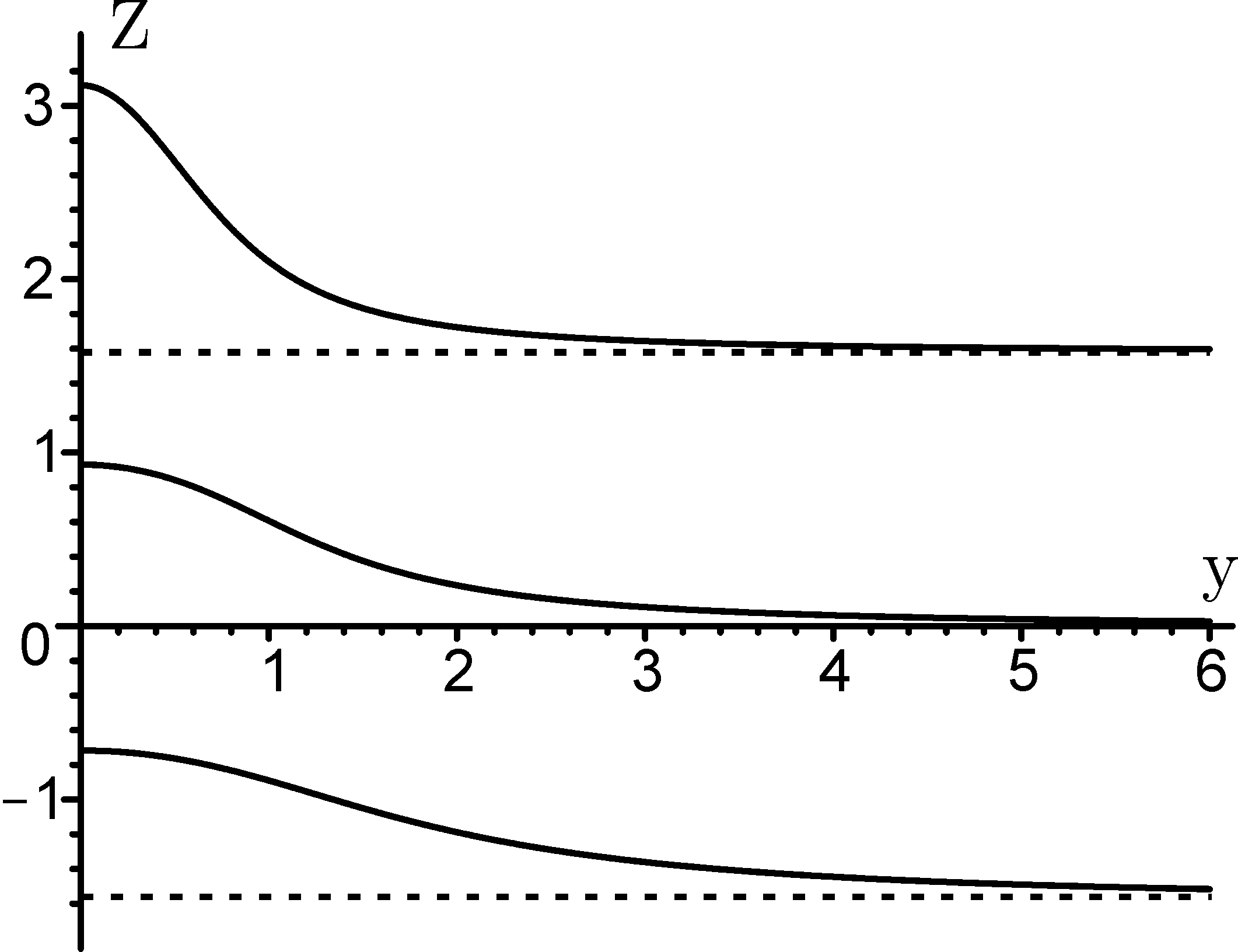}&
\includegraphics[height=4.0cm]{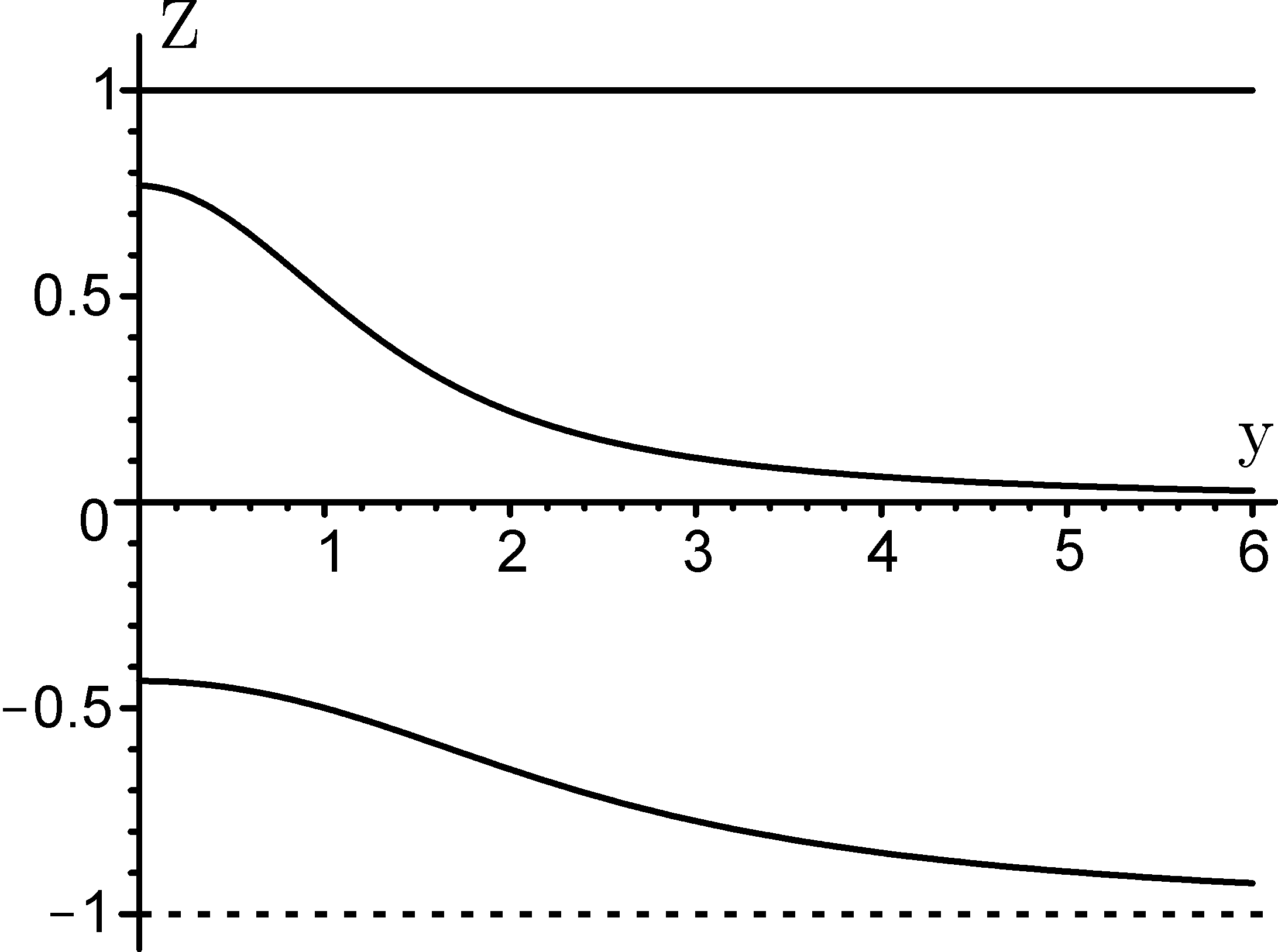}&
\includegraphics[height=4.0cm]{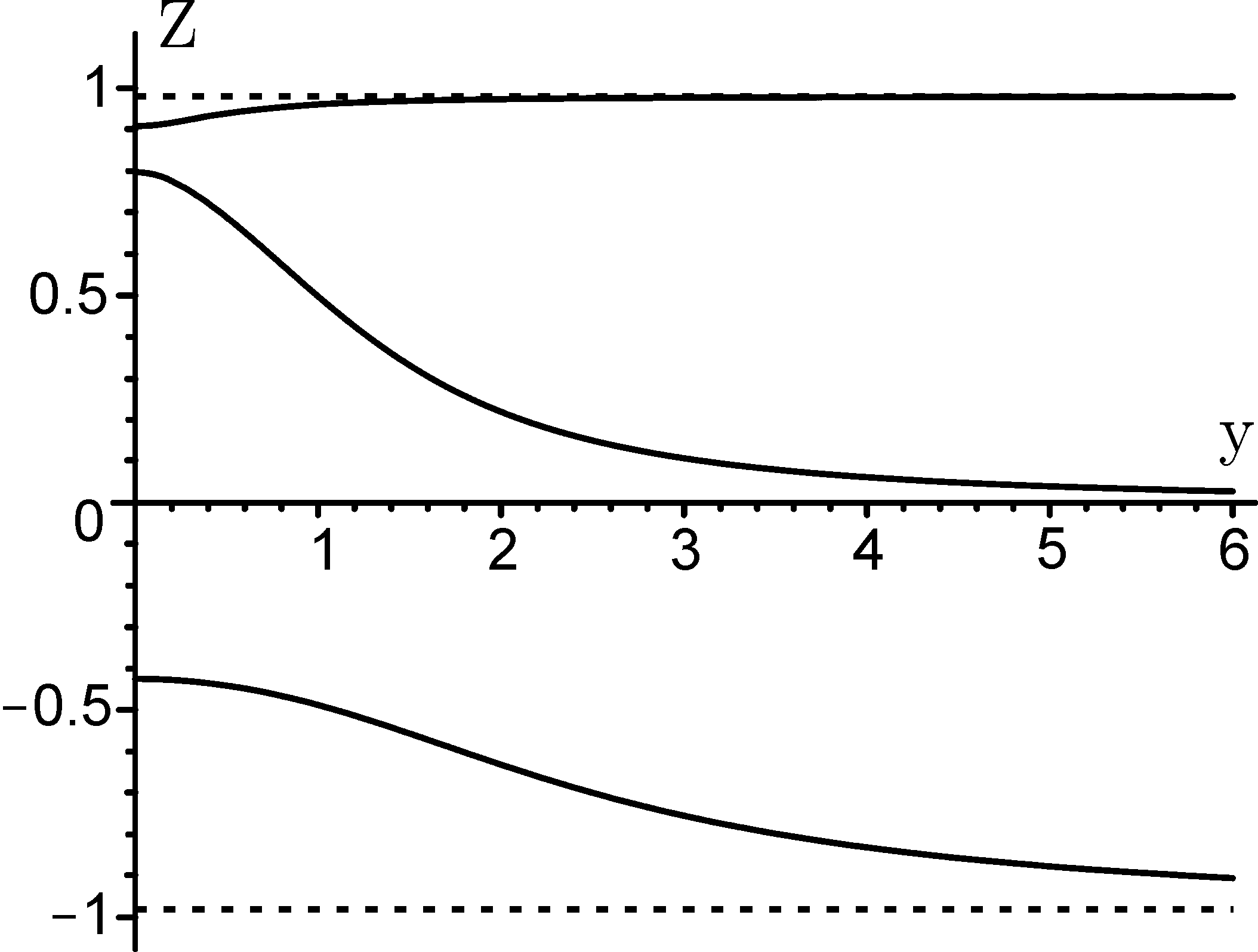}\cr
\small a) $0<a<1$ & \small b) $a=1$ & \small c) $1<a<256/243$ \cr}
\vspace{0.6cm}
\halign{\qquad\qquad\qquad\qquad\hfil#\hfil&\qquad\qquad\hfil#\hfil\cr
\includegraphics[height=4.0cm]{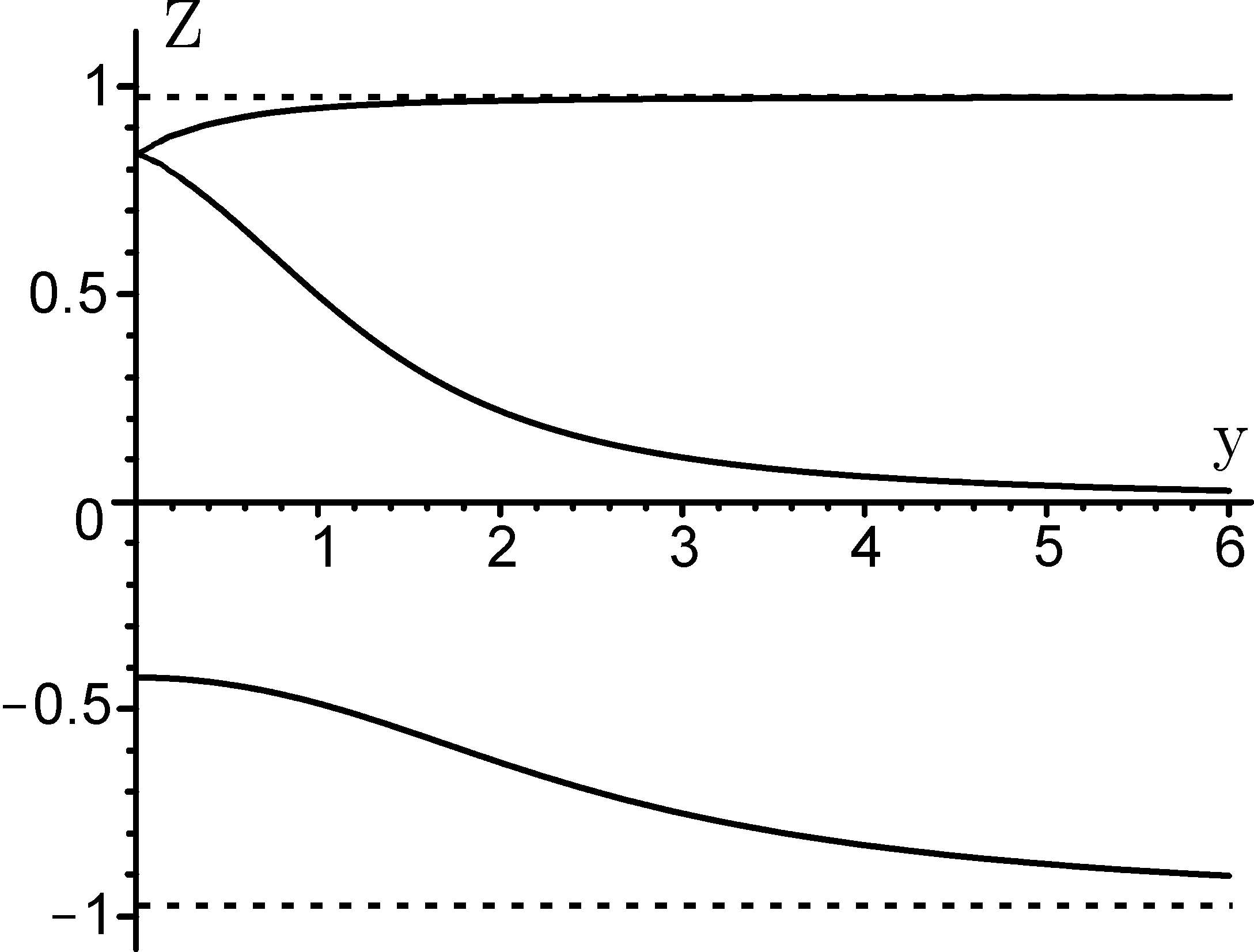}&
\includegraphics[height=4.0cm]{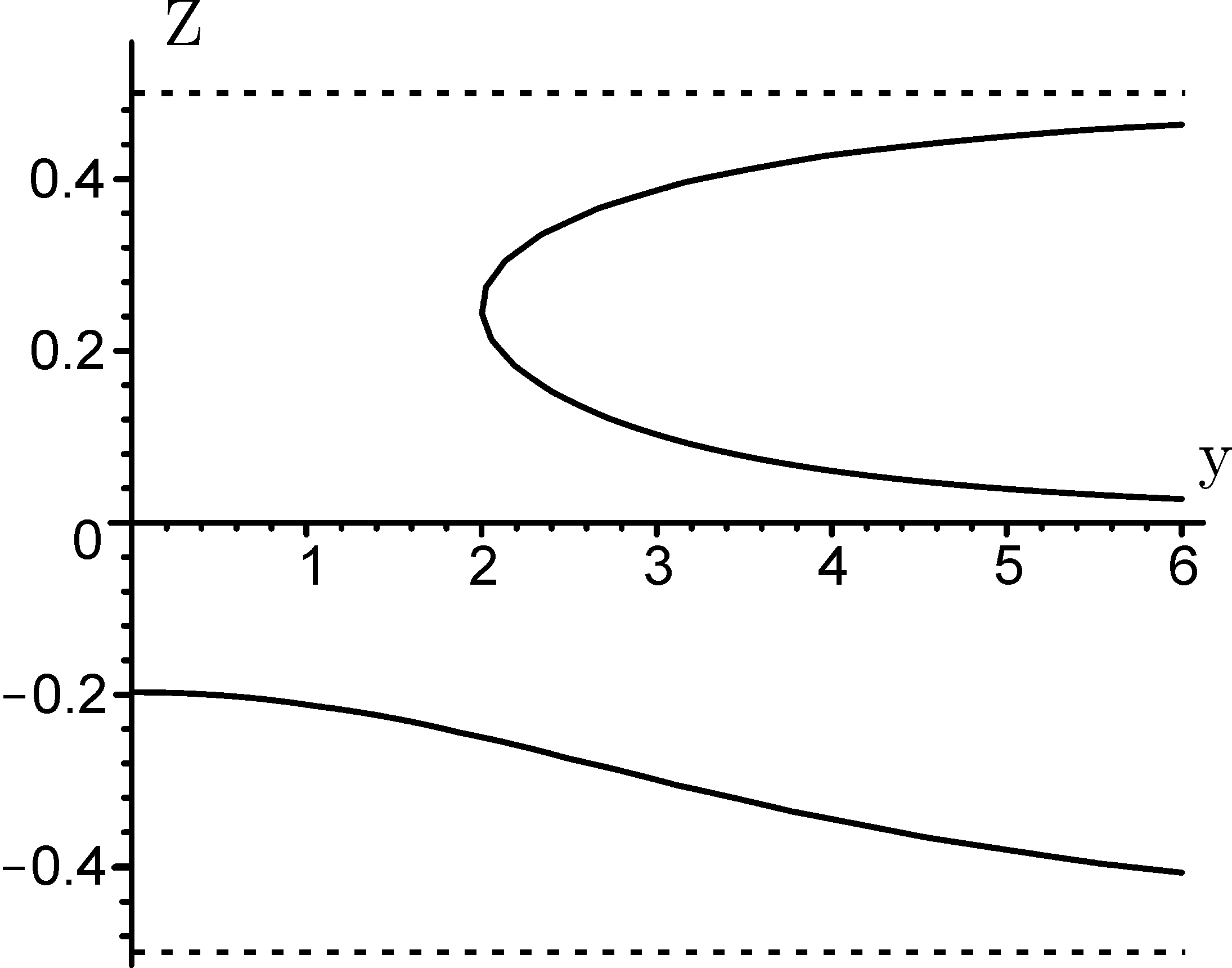}\cr
\small d) $a=256/243$ & \small e) $a>256/243$\cr}\caption{Plots of
the rescaled electric field $Z(y)$ displaying the behavior of the
Coulombian branch and the two non-Coulombian branches for $a>0$.
The interval $0<a<+\infty$ is split into three regions with
respect to two specific values of $a$, $a=1$ and $a=256/243$. The
curve $Z(y)$ is also drawn for the cases $a=1$ and $a=256/243$.
Dashed lines indicate upper ($Z=1/\sqrt{a}$) and lower
($Z=-1/\sqrt{a}$) horizontal asymptotes. (a) When $0<a<1$,
$Z_{+}(y,a)>1/\sqrt{a}$. (b) When $a=1$, the second non-Coulombian
branch degenerates into a straight line $Z_{+}(y,1)=1$. (c) When
$1<a<256/243$, $Z_{+}(y,a)<1/\sqrt{a}$. (d) When $a=256/243$, the
second non-Coulombian branch has a common point with the
Coulombian one at $y=0$. (e) When $a>256/243$, these branches have
a common point at $y>0$.} \label{posa}
\end{figure}

Now we study the $a>0$ cases. The Coulombian solution for the
electric field is the basic solution. Indeed, there are three
branches of solutions to Eq.~(\ref{E}) with $a>0$. One of them
$Z_*(y,a)$ can be identified as the Coulombian branch, the other
two can be indicated as the first and second non-Coulombian
solutions, $Z_{-}(y,a)$ and $Z_{+}(y,a)$, respectively. Let us
study first the Coulombian solution. From Eq.~(\ref{E}) this
solution is
\begin{equation}\label{Ecoul}
      Z^{-1}_{*}(y,a) =\frac{y^2}{3}+\sqrt{\frac{16a}{3}
\left[1+\frac{y^4}{12a}\right]}\,
      \cos\left\{\frac{1}{3}\arccos\left[\sqrt{\frac{243a}{256}}
\,\frac{\left(\frac{2y^6}{81a^2}+\frac{y^2}{9a}-1\right)}
      {\left(1+\frac{y^4}{12a}\right)^{3/2}}\right]\right\} \,.
\end{equation}
Clearly, $Z_{*}(y,a) \to y^{-2}$, when $y \to \infty$ for
arbitrary $a$, thus, this solution can indeed be referred to as
the Coulombian branch. This solution remains real when the modulus
of the arccosine argument does not exceed unity, i.e., $y$ belongs
to the interval given by the inequality
\begin{equation}
\sqrt{\frac{243a}{256}}\,{\left|\frac{2y^6}{81a^2}+
\frac{y^2}{9a}-1\right|}
      \leq {\left(1+\frac{y^4}{12a}\right)^{3/2}} \,.
\end{equation}
In particular, when $a \leq 256/243$, then this interval is
$0\leq y< \infty$. At $y=0$ this solution is not singular, since
\begin{equation}\label{E00}
      Z_{*}(0,a)=\frac{1}{4}\sqrt{\frac{3}{a}}\,
\cos^{-1}\left\{\frac13\arccos\left(-\sqrt{\frac{243a}{256}}
\right)\right\}\,,\quad
      0<a\leq\frac{256}{243}\,.
\end{equation}
The function $Z_{*}(0,a)$ as a function of the guiding parameter
$a$ has a minimum at $a=8/9$, the minimal value being $Z(0)_{{\rm
min}}=3/4$. The behavior of the Coulombian branch of the solution
for $E(r)$ can be illustrated in a figure. In Fig.~\ref{posa}a-e
we plot the rescaled electric field $Z(y)$ as a function of the
rescaled radius $y$, for different values of $a$. The Coulombian
branch $Z_*(y,a)$ is represented by the middle line.

The second solution $Z_{-}(y,a)$ of the cubic equation (\ref{E})
can be represented in the form
\begin{equation}\label{Edown}
      Z^{-1}_{-}(y,a)=\frac{y^2}{3}-\sqrt{\frac{16a}{3}
\left[1+\frac{y^4}{12a}\right]}\,
      \cos\left\{\frac{1}{3}\arccos
\left[\sqrt{\frac{243a}{256}}\,\frac{
\left(1-\frac{y^2}{9a}-\frac{2y^6}{81a^2}\right)}
      {\left(1+\frac{y^4}{12a}\right)^{3/2}}\right]\right\}\,.
\end{equation}
At $y\to \infty$ this curve tends to the horizontal asymptote
$Z=-1/\sqrt{a}$, and in this sense it is non-Coulombian, we call
it the first non-Coulombian solution. The solution is real for
arbitrary (positive) value of the guiding parameter $a$, see the
lower curves of the Fig.~\ref{posa}a-e. At the point $y=0$, the
center, this solution is also nonsingular
\begin{equation}
\begin{array}{ll}
\displaystyle
Z_{-}(0)=-\frac{1}{4}\sqrt{\frac3a}\,\cos^{-1}\left\{\frac13
\arccos\left(\sqrt{\frac{243a}{256}}\right)\right\}, &
      \quad a < 256/243\,,\\
\vphantom{\biggl( } Z_{-}(0)=-27/64 \,, & \quad a=256/243 \,,\\
\displaystyle Z_{-}(0)=-\frac{1}{4}\sqrt{\frac3a}\,\cosh^{-1}
\left\{\frac13\arccosh\left(\sqrt{\frac{243a}{256}}\right)\right\},
& \quad a > 256/243\,.
\end{array}
\end{equation}

The third solution of the cubic equation (\ref{E}) is of the form
\begin{equation}\label{Eup}
      Z^{-1}_{+}(y,a) = \frac{y^2}{3}+\sqrt{\frac{16a}{3}
\left[1+\frac{y^4}{12a}\right]}\,
      \sin\left\{\frac{1}{3}\arcsin\left[\sqrt{\frac{243a}{256}}
\,\frac{\left(1-\frac{y^2}{9a}-\frac{2y^6}{81a^2}\right)}
      {\left(1+\frac{y^4}{12a}\right)^{3/2}}\right]\right\} \,.
\end{equation}
For $y\to \infty$ this curve tends to the horizontal asymptote
$Z=1/\sqrt{a}$, thus the solution is also non-Coulombian, it is
the second non-Coulombian solution. Nevertheless, now the curve
behaves analogously to the Coulombian one, for instance, the
solution is real for $0\leq y<+\infty$ when $a\leq 256/243$, see
the upper curves of the Figs.~\ref{posa}a-e. The value at the
center is
\begin{equation}
Z_{+}(0,a)=\frac{1}{4}\sqrt{\frac3a}\,\sin^{-1}
\left\{\frac13\arcsin\left(\sqrt{\frac{243a}{256}}\right)\right\}
\,, \quad a\leq \frac{256}{243}\,.
\end{equation}
Clearly, $Z_{*}(0)=Z_{+}(0) = 27/32$, when $a= \frac{256}{243}$,
i.e., the curves $Z_{*}$ and $Z_{+}$ contact at $y=0$ at this
value of the guiding parameter $a$, see Fig.~\ref{posa}e.  When $a
> 256/243$ the curves $Z_{*}$ and $Z_{+}$ contact at $y > 0$, the
inverse function $y(Z)$ being smooth in the vicinity of this
point.  Thus, when $a>\frac{256}{243}$ one has that $y_{{\rm
min}}>0$, and there is the possibility of finding a throat for a
traversable wormhole.

Let us also mention that $Z_{+}(0,a)>Z_{+}(\infty,a)$ when $a<1$,
and $Z_{+}(0,a)<Z_{+}(\infty,a)$ when $a>1$ (see
Figs.~\ref{posa}\,a,\,c).  In this connection we would like to
make the following remarks: The cubic equation \Ref{E} differs
from the analogous equation discussed in \cite{BBL08} [see
\Ref{0E1}] by the coefficients in front of $Z^3$ and $Z^2$. This
leads to the following differences: {\it (i)} In contrast to
\cite{BBL08} the curves $Z_{*}$ and $Z_{+}$ do not intersect at
$a=1$.  {\it (ii)} There are no bubbles in the vicinity of the
center in the framework of this new model.  {\it (iii)} The
branches $Z_{+}$ and $Z_{*}$ can contact at $y=0$, it is possible,
when $a=256/243$.

\section{Exact solutions: Solutions with a center}\label{III}

\subsection{Behavior of the function $\sigma(\rho)$}

Based on the study of the cubic equation for the electric field,
we consider the solutions for the metric functions $\sigma(r)$ and
$N(r)$. We focus here on the solutions with center, i.e., on the
cases, when the point $Y=0$ is accessible. The solution with
center for the nonminimal electric field with Coulombian asymptote
exists, when $0<a\leq 256/243$ (see Figs.~\ref{posa}a-d). Of
course, solutions with center with non-Coulombian asymptotes exist
for other $a$.

Let us consider the behavior of the function $\sigma(\rho)$ for
the case $0<a\leq 256/243$. We can put now $Y=r$ and,
respectively, $y=\rho=\frac{r}{r_Q}$, keeping in mind the
asymptotic behavior (\ref{asympCond}). Then the formula \Ref{s}
takes the form
\begin{equation}
      \sigma=1-2a\left(Z^2+\rho Z\,\frac{dZ}{d\rho}\right)\,.
\end{equation}
According to Fig.~\ref{posa}a-d, $Z(0)$ and $\frac{dZ(0)}{d\rho}$
are finite and the value of the function $\sigma(\rho)$ at the
origin reads
\begin{equation}\label{sig0}
      \sigma(0,a)=1-2aZ^2(0,a) \,.
\end{equation}
Substituting \Ref{E00} into \Ref{sig0} we obtain for $0<a\leq
\frac{256}{243}$
\begin{equation}\label{sig00}
      \sigma(0,a)=1-\frac{3}{8}\cos^{-2}\left\{\frac{1}{3}
\arccos\left(-\sqrt{\frac{243a}{256}}\right)\right\}\,.
\end{equation}
The value $\sigma(0,a)$ as a function of the guiding parameter $a$
is steadily decreasing with increasing $a$. When $a\to 0$, this
function tends to $1/2$, when $a=256/243$ then $\sigma(0,a)=-1/2$,
and when $a=8/9$ the function \Ref{sig00} takes the value zero.
When $8/9\leq a\leq 256/243$, the function $\sigma(0,a)$ is
nonpositive. Thus taking into account the asymptotic condition
$\sigma(\infty,a)=1>0$ there is at least one root $\rho_s\geq 0$
of the equation $\sigma(\rho)=0$. When $a<8/9$, the value
$\sigma(0,a)$ is positive.

\subsection{Exact solution for $a=1$}

When $a=1$ Eq.~(\ref{E}) admits the solution $Z=1$, which it is
one of the curves from the family $Z_{+}(y,a)$, namely,
$Z_{+}(y,1)$. Thus, in this case the cubic equation (\ref{E})
splits, so that the Coulombian solution $Z_{*}$ and the first
non-Coulombian solution $Z_{-}$ can be obtained as solutions of
the corresponding quadratic equations. This procedure yields,
\begin{equation}\label{Eexam2}
      Z_{*}(y,1)=\frac{\sqrt{13+2y^2+y^4}+1-y^2}{2(3+y^2)}\,,\quad
      Z_{*}(0,1)=\frac{1 + \sqrt{13}}{6} \,,
\end{equation}
\begin{equation}\label{Eexam3}
      Z_{-}(y,1)=\frac{-\sqrt{13+2y^2+y^4}+1-y^2}{2(3+y^2)}\,,\quad
      Z_{-}(0,1)=\frac{1- \sqrt{13}}{6} \,.
\end{equation}

\begin{figure}[h]
\includegraphics[height=4cm]{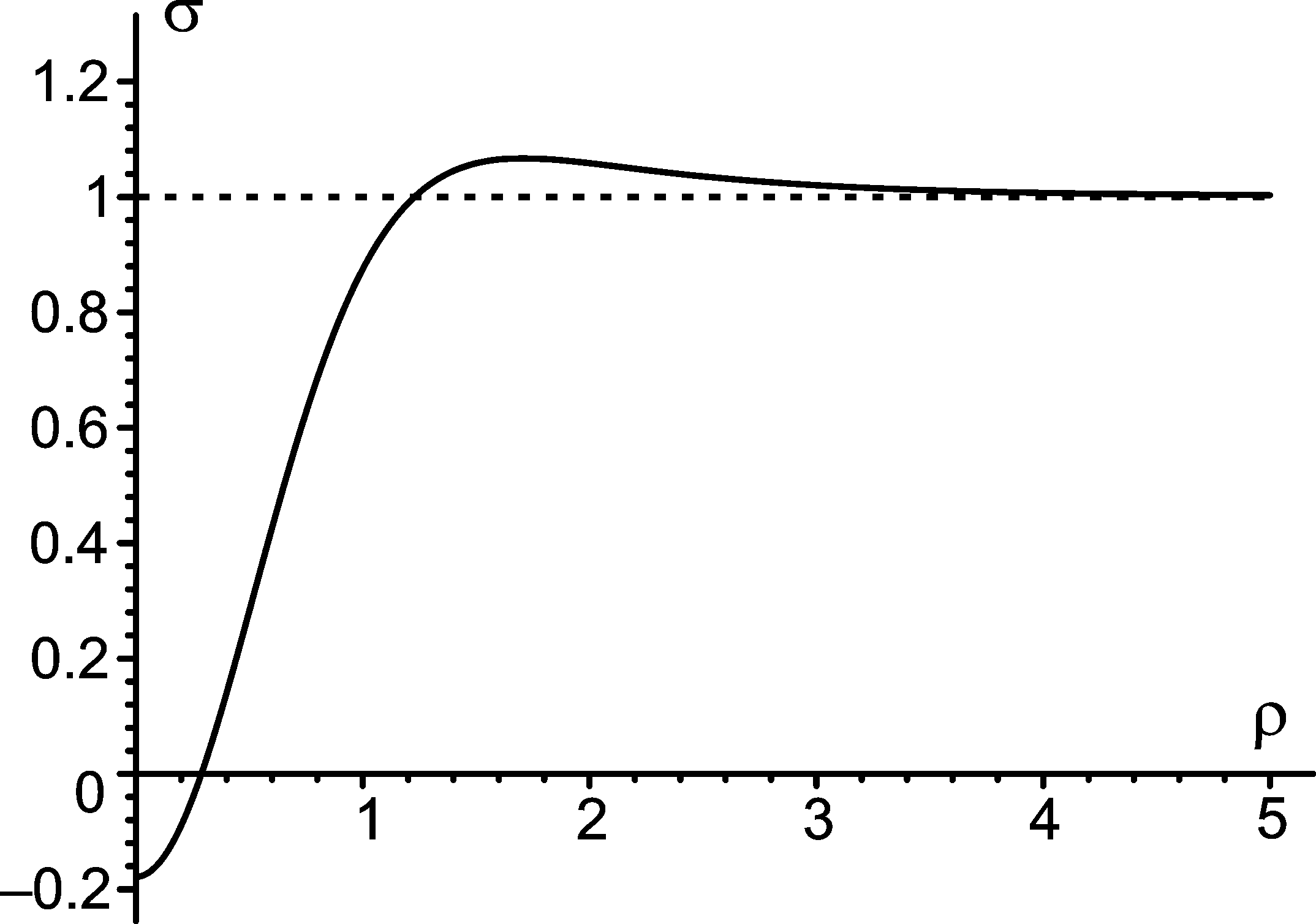}
\caption{Plot of the metric function $\sigma(\rho)$ at $a=1$. This
function has only one root $\rho_s=0.289$.} \label{fig3}
\end{figure}
\begin{figure}[t]
\halign{\qquad\hfil#\hfil&\qquad\hfil#\hfil&\qquad\hfil#\hfil\cr
\includegraphics[height=3.5cm]{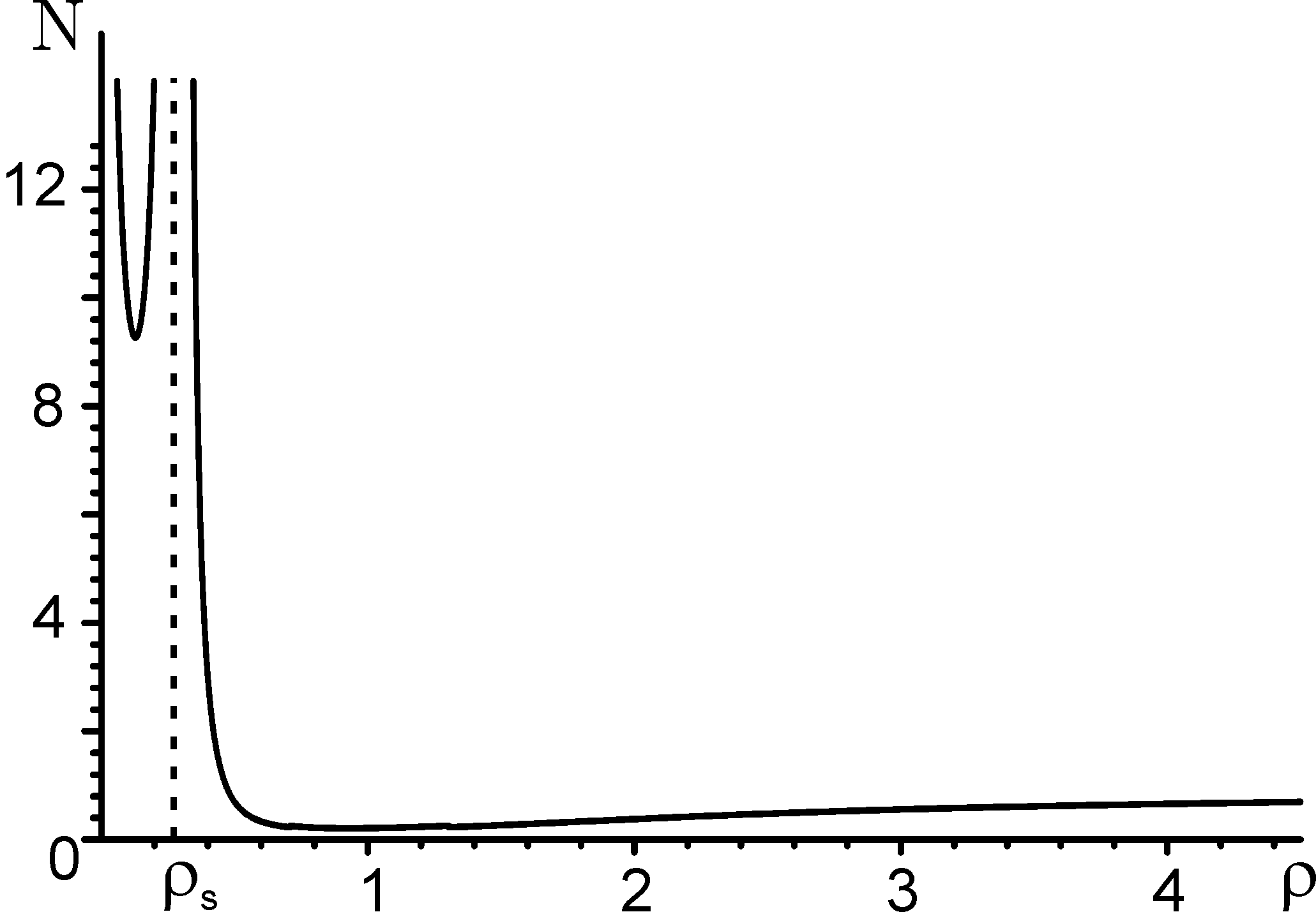}&
\includegraphics[height=3.5cm]{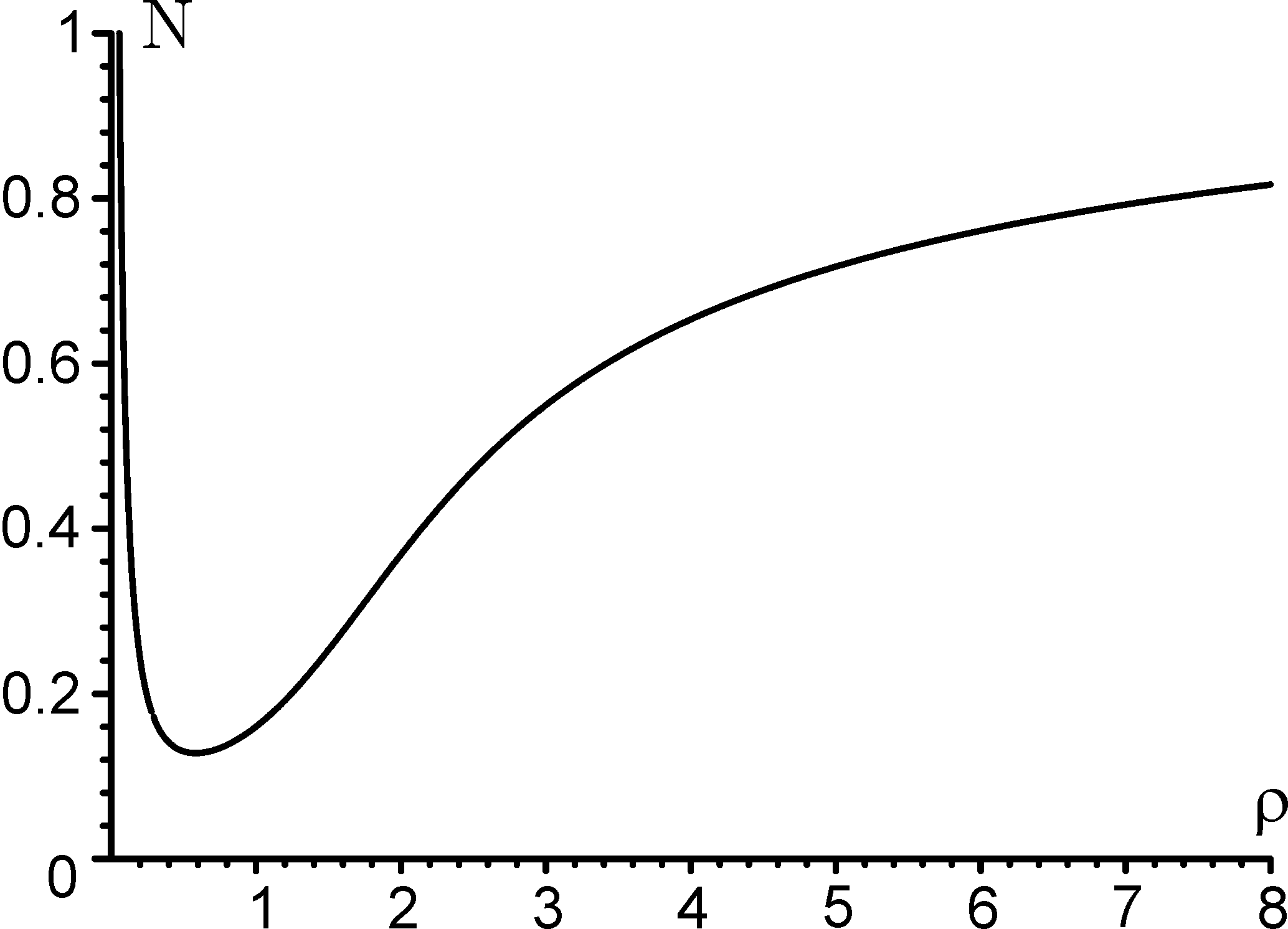}&
\includegraphics[height=3.5cm]{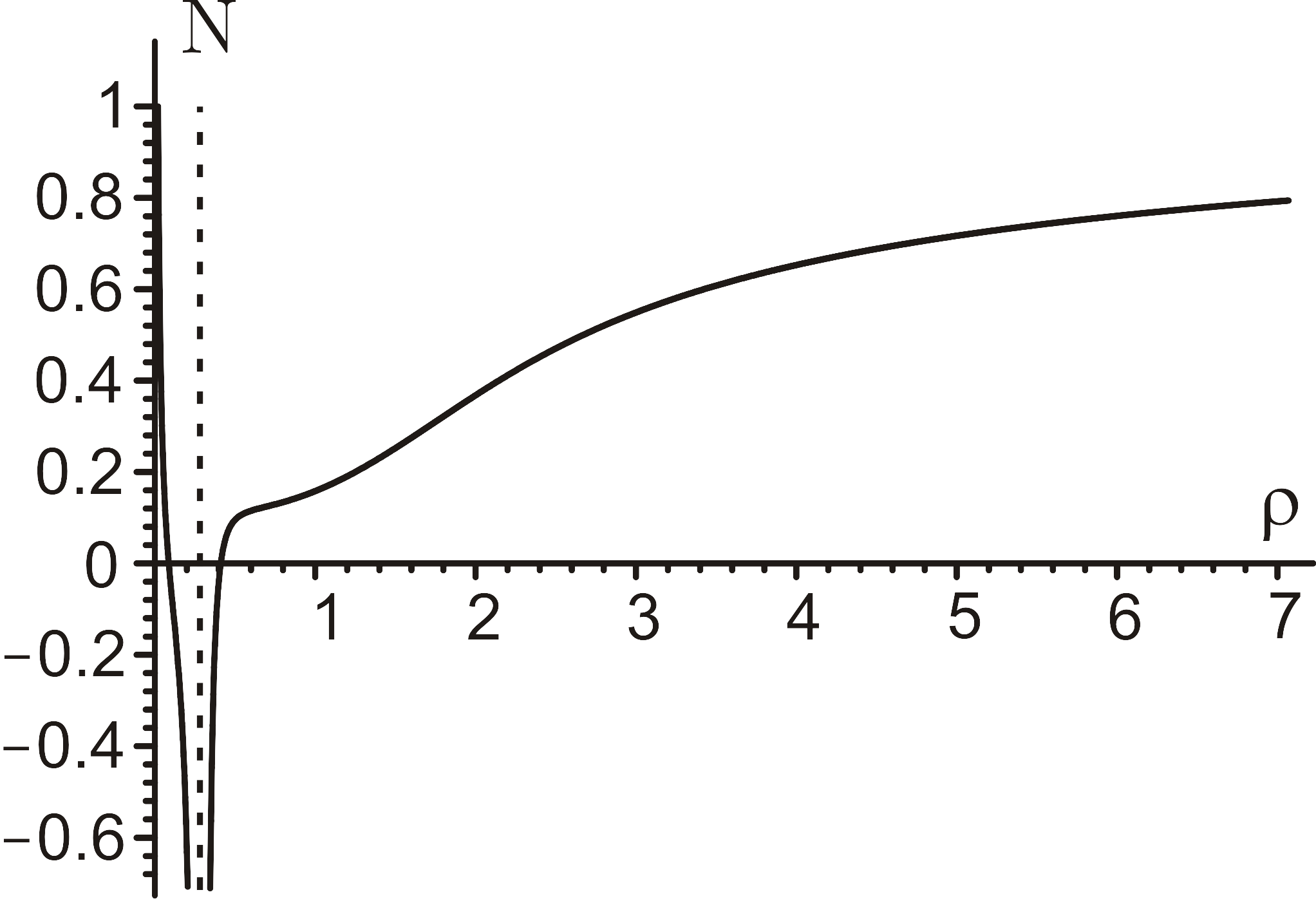}\cr
\small a) $0<m<m^*$& b) $m=m^*=0.7913$ & c) $m^*<m<m_0$ \cr}
\vspace{0.6cm}%
\halign{\qquad\qquad\qquad\qquad\hfil#\hfil&\qquad\hfil#\hfil\cr
\includegraphics[height=3.5cm]{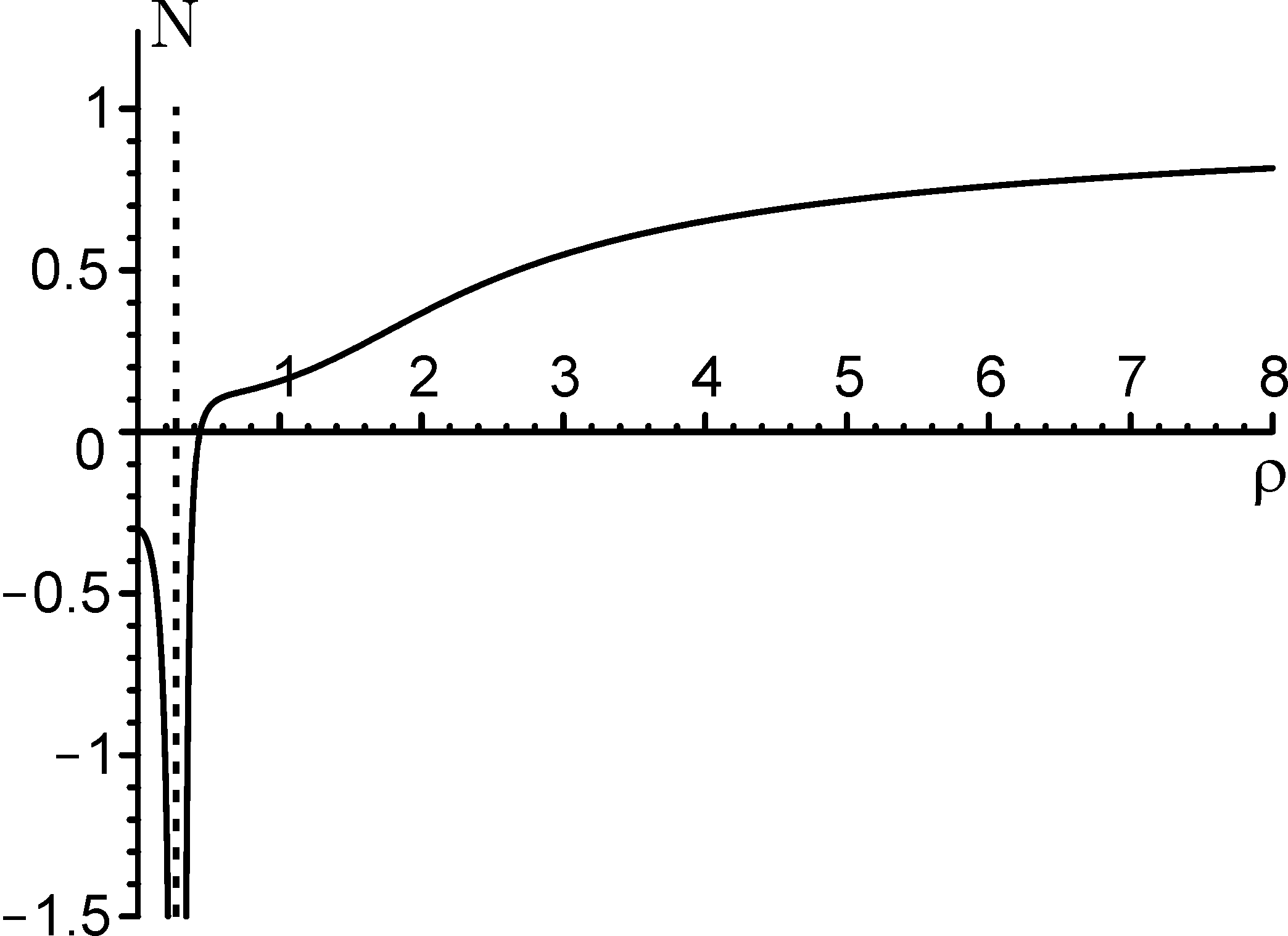}&
\includegraphics[height=3.5cm]{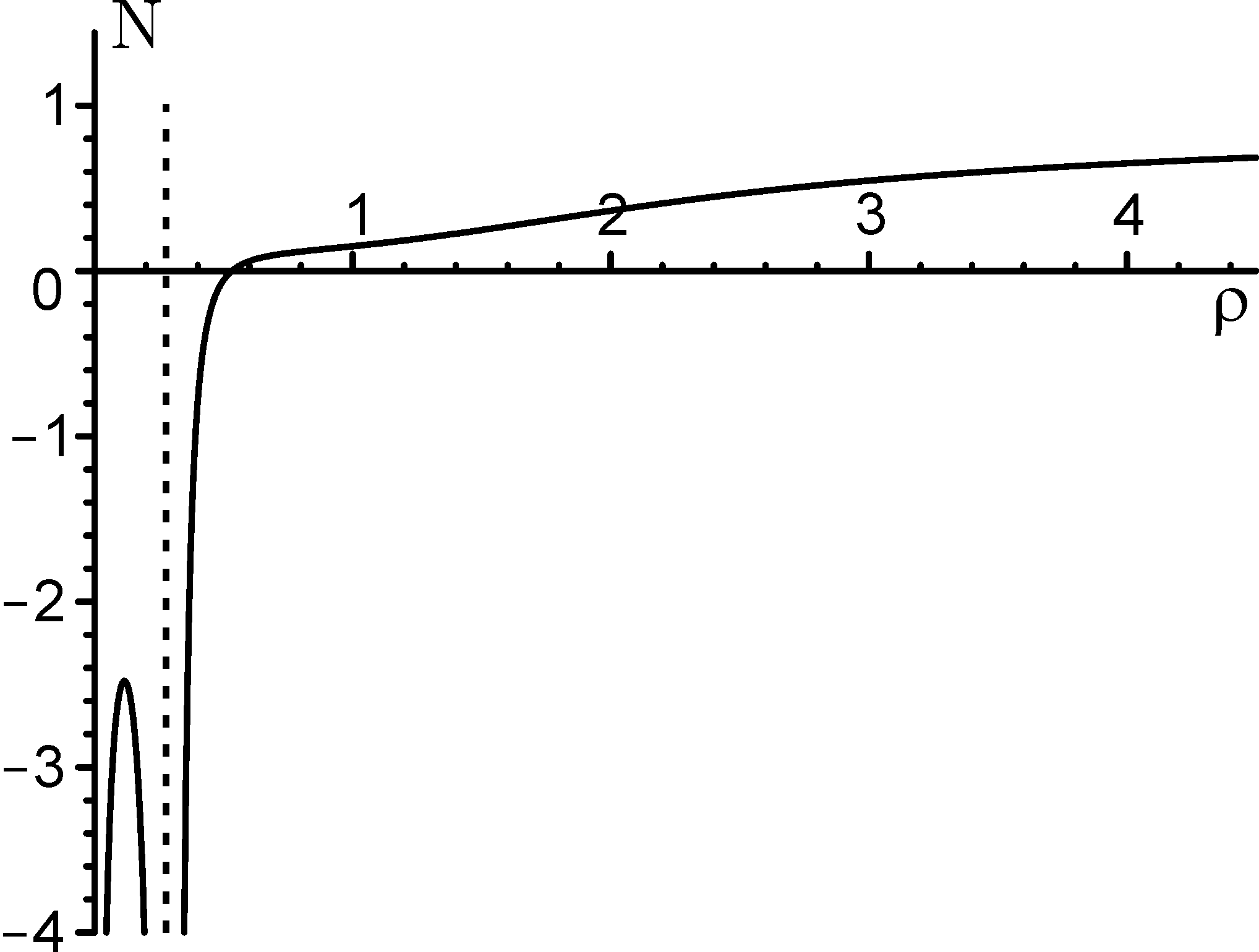}\cr
\small d) $m=m_0=0.7924$& e) $m>m_0$ \cr} \caption{Plots of the
function $N(\rho)$ for $a=1$ for various values of the rescaled
asymptotic mass $m=M/r_Q$. We split the interval $0<m<+\infty$
into three regions with respect to two specific values of mass:
$m=m^*=0.7913$ and $m=m_0=0.7924$. All curves except (b) possess a
vertical asymptote at $\rho=\rho_s=0.289$. When $m=m_0$ [plot (d)]
$N(0)$ is finite, contrary to the other cases.}\label{fig5}
\end{figure}
\begin{figure}[t]
\halign{\hfil#\hfil&\qquad\hfil#\hfil&\qquad\hfil#\hfil\cr
\includegraphics[height=3.8cm]{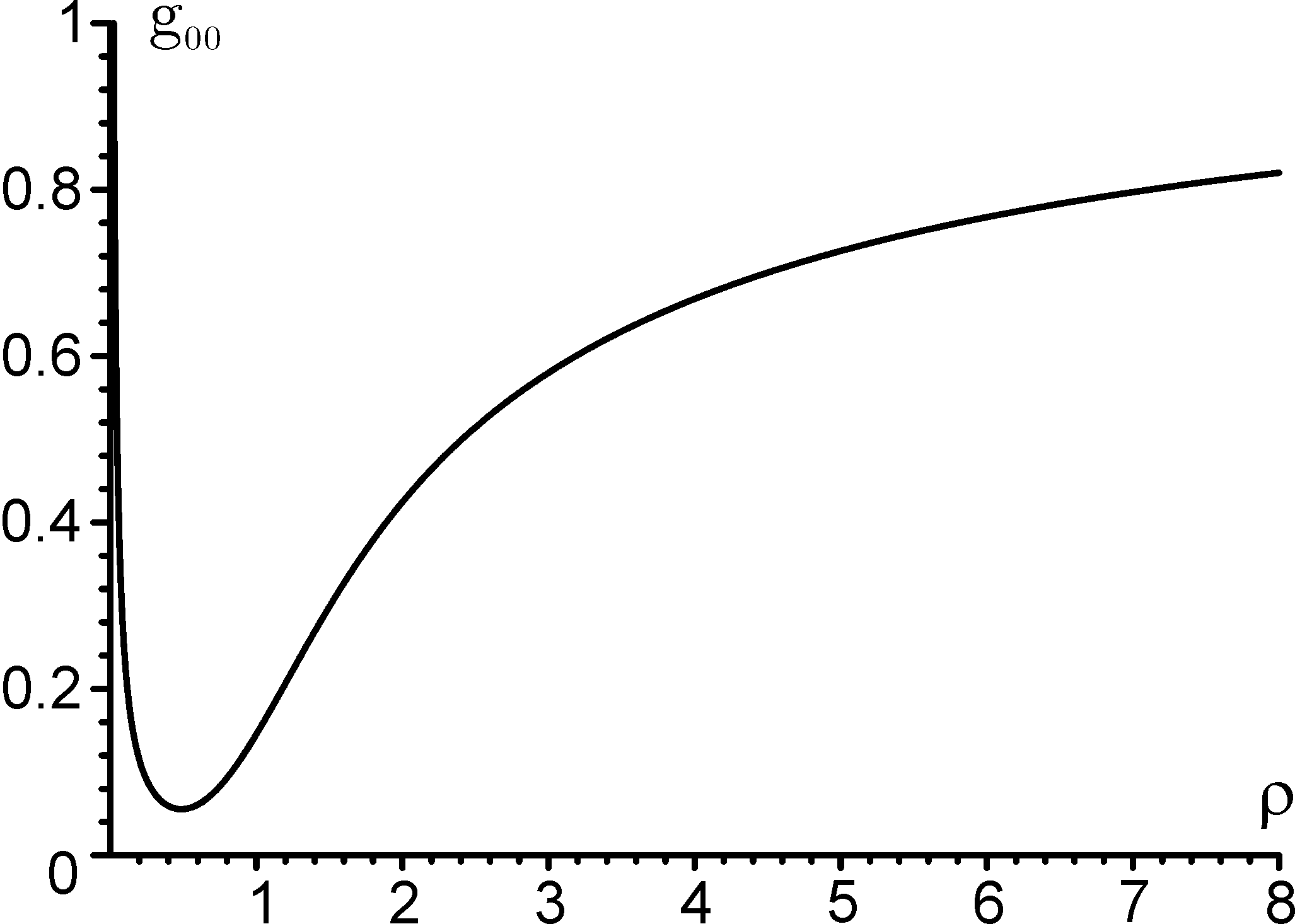}&
\includegraphics[height=3.8cm]{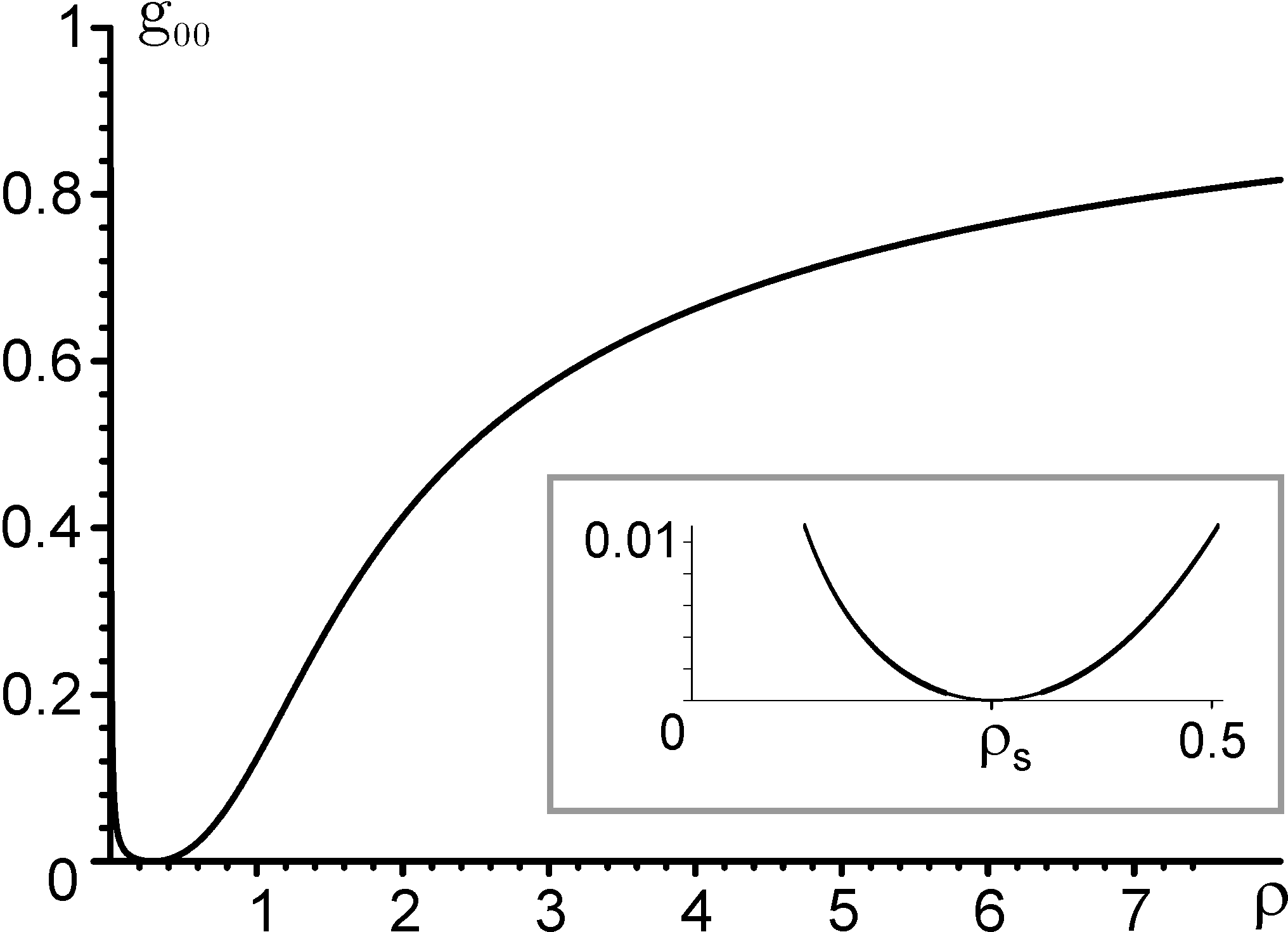}&
\includegraphics[height=3.8cm]{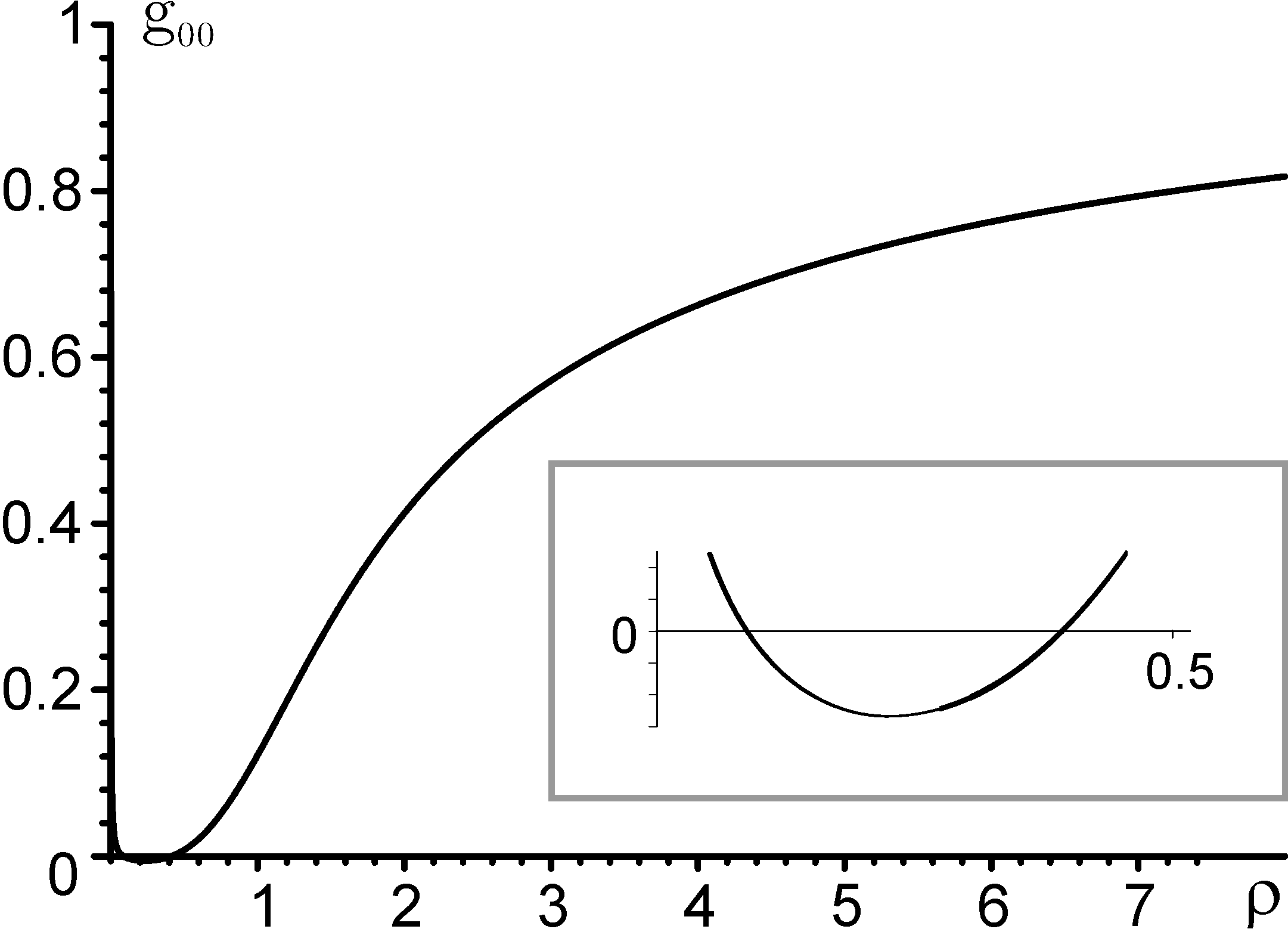}\cr
\small a) $0<m<m^*$& b) $m=m^*=0.7913$ & c) $m^*<m<m_0$ \cr}
\vspace{0.6cm}%
\halign{\qquad\qquad\qquad\qquad\hfil#\hfil&\qquad\hfil#\hfil\cr
\includegraphics[height=3.8cm]{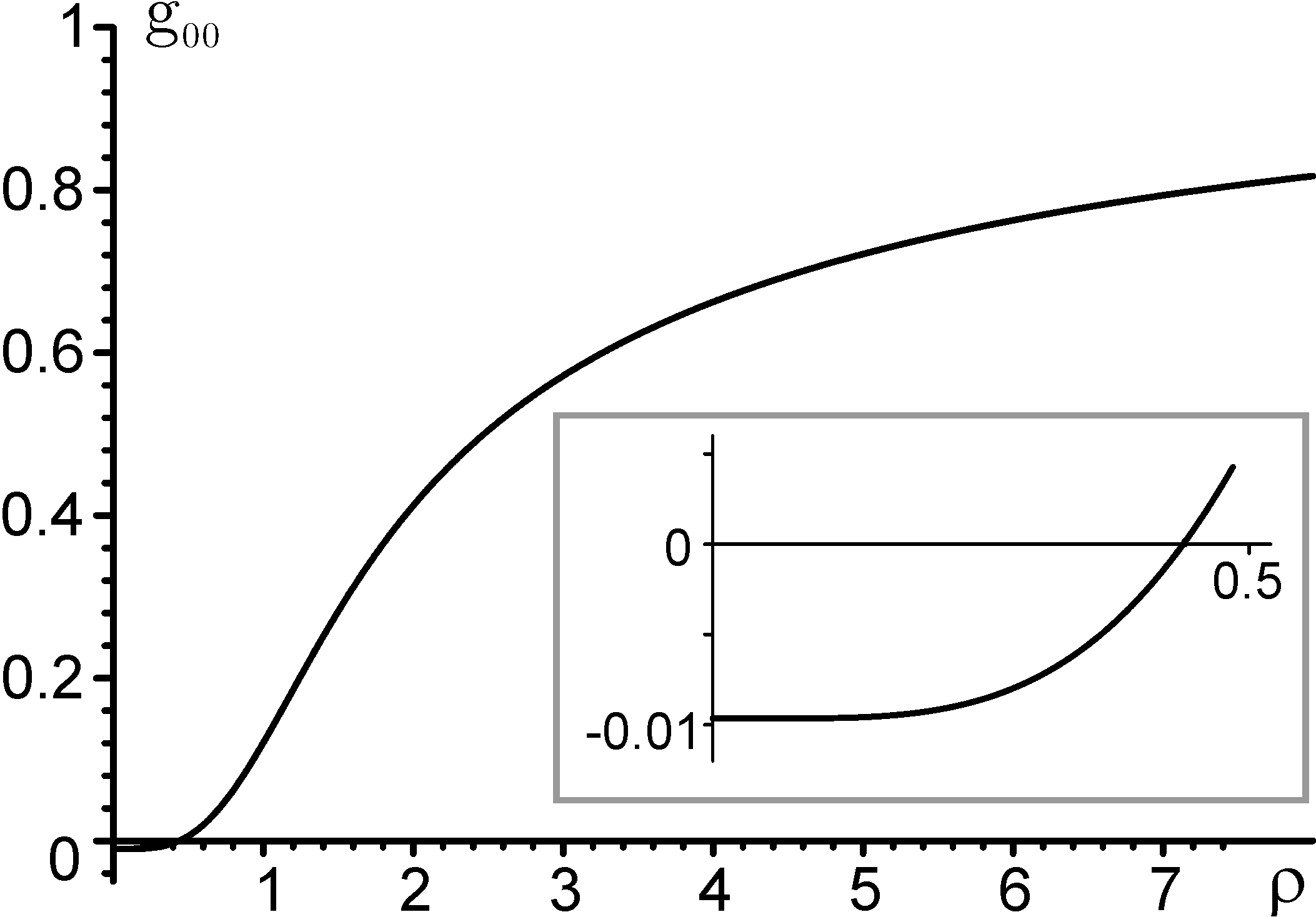}&
\includegraphics[height=3.8cm]{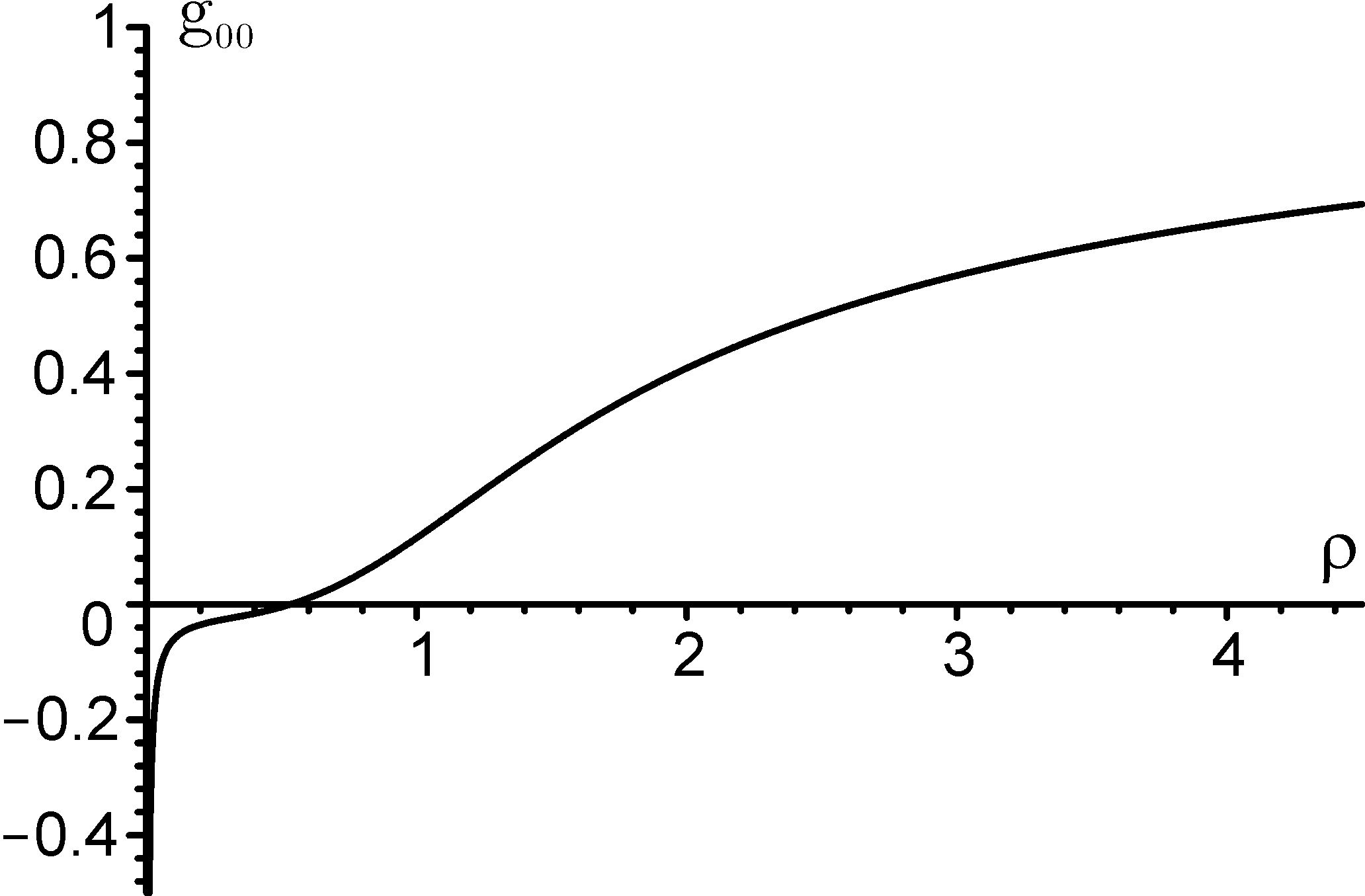}\cr
\small d) $m=m_0=0.7924$& e) $m>m_0$ \cr} \caption{Plots of the
function $g_{00}(\rho)=\sigma^2N$ for $a=1$ for various values of
the rescaled asymptotic mass $m=M/r_Q$. We split the interval
$0<m<+\infty$ into three regions with respect to two specific
values of mass: $m=m^*=0.7913$ and $m=m_0=0.7924$. (a) When
$m<m^{*}$, the function $g_{00}$ is positive and tends to infinity
at origin. (b) When $m=m^*$, the curve touches the horizontal axis
at $\rho=\rho_s=0.289$. (c) When $m^*<m<m_0$, the curve intersects
the horizontal axis at two different points; the function $g_{00}$
is negative between these points. (d) When $m=m_0$, the curve
intersects the horizontal axis at one point; the function $g_{00}$
tends to a finite negative value at origin. (e) When $m>m_0$, the
function $g_{00}$ tends to negative infinity at
origin.}\label{fig4}
\end{figure}

We now study the behavior of the functions $\sigma(\rho)$ and
$N(\rho)$ for this model with $a=1$ which is of interest for
illustrative purposes. In this case the Coulombian branch of the
electric field is described by (\ref{Eexam2}), thus the metric
function $\sigma(\rho)$ is of the form
\begin{equation}
\sigma(\rho)=\frac{(6+34\rho^2)\sqrt{13+
2\rho^2+\rho^4}-39+82\rho^2+10\rho^4+
10\rho^6+\rho^8}{(3+\rho^2)^3\sqrt{13+2\rho^2+\rho^4}} \,.
\end{equation}
The plot of the function $\sigma(\rho)$ is represented in
Fig.~\ref{fig3}. At the origin
$\sigma(0,1)=-(\sqrt{13}-2)/9\approx -0.178$ is negative. This
function has only one root $\rho_s=0.289$. At $\rho=1.706$ it
reaches a maximum, $\sigma_{\rm max}=1.067$, and then tends to one
asymptotically.

The constant in (\ref{N}) is connected with the asymptotic mass
$M$ defined by $M{=}\frac{1}{2}\lim\limits_{r \to\infty}[r
(1{-}g_{00})]$, where $g_{00}=\sigma^2N$. Taking into account this
condition one obtains
\begin{gather}
      g_{00}=1-\frac{2m-J(\rho)}{\rho}\,,\qquad N=
\frac{1}{\sigma^2}\left[1-\frac{2m-J(\rho)}{\rho}\right]\,,\nonumber\\
      J(\rho)=\int\limits_\rho^\infty d\rho
      \left[1-\frac{\sigma}{2(3+\rho^2)^2}
\left(19+23\rho^2+5\rho^4+\rho^6-(5+2\rho^2+\rho^4)
\sqrt{13+2\rho^2+\rho^4}\right)\right]\,,
\end{gather}
where $m\equiv M/r_Q$. The value of the integral $J(\rho)$ at
$\rho=0$ is finite and equal to $J(0)=1.5848$. Depending on the
value of mass $m$ there are three variants of the behavior of the
functions $g_{00}(\rho)$ and $N(\rho)$ in the vicinity of
$\rho=0$. They are: \noindent {\it (i)}
$m<m_0=\frac{1}{2}J(0)=0.7924$. \noindent The function $g_{00}$
tends to positive infinity, when $\rho\to 0$ [see
Fig.~\ref{fig4}\,a-c]. \noindent {\it (ii)} $m=m_0$. \noindent The
function $g_{00}$ takes the negative finite value
$g_{00}(0)=-0.0096$ [see Fig.~\ref{fig4}\,d]. \noindent {\it
(iii)} $m>m_0$. \noindent The function $g_{00}$ tends to negative
infinity, when $\rho\to 0$ [see Fig.~\ref{fig4}\,e]. Plots of the
function $N(\rho)$ and of the function $g_{00}(\rho)=\sigma^2N$
for $a=1$ are given in Figs.~\ref{fig5} and \ref{fig4},
respectively.

In addition to the special value of the mass $m=m_0$, there exists
another special value $m^*=0.7913$, which can be obtained as
follows. The function $\sigma(\rho)$ takes zero value at
$\rho=\rho_s$ (see Fig.~\ref{fig3}), and $N(\rho)=g_{00}/\sigma^2$
generally becomes infinite at this point. Nevertheless, the
possibility exists to make $N(\rho_s)$ finite, when $\rho_s$ is a
double root of the function $g_{00}$. This situation takes place,
when $m=m^*$, see Fig.~\ref{fig5}\,b for $N(\rho)$ and
Fig.~\ref{fig4}\,b for $g_{00}(\rho)$. However, for both cases the
Kretschmann scalar ${\cal K}=R_{ikmn}R^{ikmn}$ is infinite at this
point, ${\cal K} \propto (\rho-\rho_s)^{-6}$ when $m\neq m^*$ and
${\cal K} \propto (\rho-\rho_s)^{-2}$ when $m=m^*$. Hence, the
sphere with the radius $\rho=\rho_s$ is a singularity. In addition
there exists a second singularity at the origin. It is of a
Schwarzschild type, when $m\neq m_0$, and becomes a conic
singularity, when $m=m_0$, since the functions $\sigma$ and $N$
are finite at $\rho=0$, but $N(0)\neq1$. When $m> m^*$ these
singularities are hidden inside an event horizon, while for the
case $m\leq m^*$ they are naked, see Figs.~\ref{fig5} and
\ref{fig4}. This behavior differs from the situation described in
\cite{BaDeZa09} for the magnetic monopole with the same set of
nonminimal coupling constants.

%\newpage
\section{Exact solutions: Wormholes (solutions without a
center)} \label{IV}

\subsection{The expression for $Y$}\label{C}

According to Fig.~\ref{posa}e, when $a>256/243$, the value $Y=0$
is inaccessible for the Coulombian branch of the electric field.
It means that $Y(r)$ reaches a minimal value $Y_{\rm min}(a)$, for
which the Coulombian branch smoothly joins with the second
non-Coulombian one. Therefore we are in the presence of a
wormhole. The corresponding surface of minimal area $4\pi Y_{\rm
min}^2$ is its throat.

In order to describe this object let us introduce a new radial
coordinate $r$, which covers the entire range $(-\infty;+\infty)$.
Without loss of generality, we can take as usual the throat to
occur at $r=0$
\begin{equation}\label{thrCond}
      Y'(0)=0\,,\quad Y''(0)>0\,.
\end{equation}
Also we assume that positive values of $r$ are associated with the
Coulombian branch of the electric field, while the range
$(-\infty;0)$ corresponds to the non-Coulombian one we are
studying. We have to supplement the asymptotic form
(\ref{asympCond}) for the function $Y(r)$ with a condition at
negative infinity
\begin{equation}\label{asympCondwh}
     \lim_{r\to -\infty} Y(r)= +\infty \,,
\end{equation}
which correlates with Fig.~\ref{posa}e.

\subsection{Radius of the throat}

\begin{figure}[h]
\includegraphics[height=4.0cm]{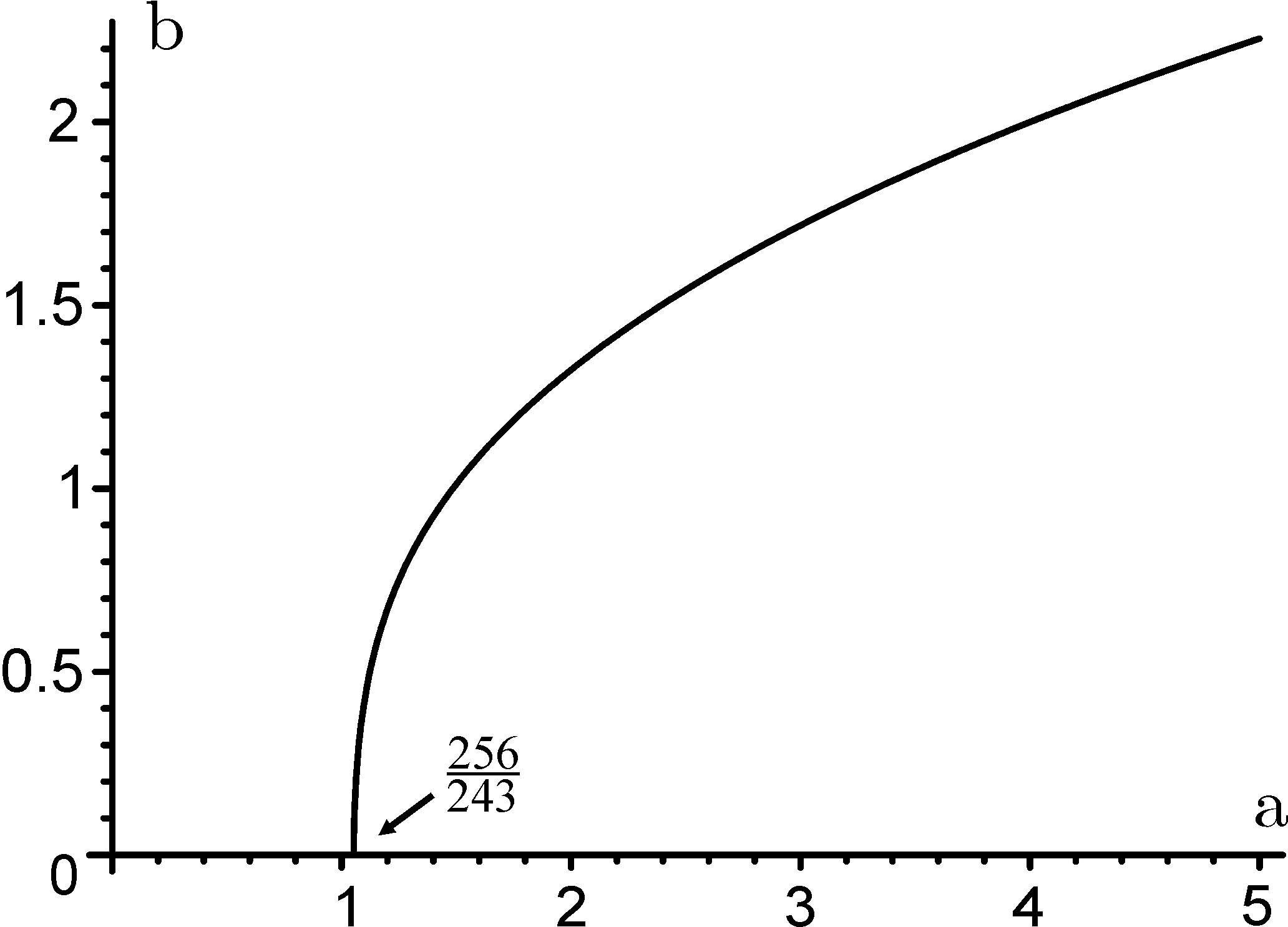}
\caption{Plot of the dimensionless radius of the wormhole throat
$b$ as a function of the parameter $a$. It starts at $a=256/243$,
i.e., if the parameter $a$ is less than this value a throat does
not exist.}\label{fig6}
\end{figure}

Further analytical progress is possible if we fix an explicit
expression for this function, satisfying both the asymptotic
condition (\ref{asympCondwh}) and the throat conditions
(\ref{thrCond}). The simplest function of such a type is
\begin{equation}\label{Yworm}
      Y(r)=\sqrt{r^2+r_Q^2b^2}\,,
\end{equation}
or, after rescaling (\ref{dimless})
\begin{equation}\label{ywh}
      y(\rho)=\frac{Y}{r_Q}=\sqrt{\rho^2+b^2}\,,
\end{equation}
where $b$ is the dimensionless radius of the throat. The value of
$b$ can be found as follows: In Eq.~(\ref{Ecoul}) the argument of
the function arc cosine takes the maximum value $+1$ corresponding
to $y\to\infty$. The minimal value $-1$ relates to $y=y_{\rm
min}=b$. Thus the value of the throat's radius satisfies the
equation
\begin{equation}\label{rth}
b^8+3ab^6+\frac{61}{4}ab^4+\frac{27}{2}a^2b^2-\frac{a^2}{4}(243a-256)=0\,.
\end{equation}
Let us mention that the argument of the arc sine in
Eq.~(\ref{Eup}) takes the value $+1$, and $Z_{*}(b,a)=Z_{+}(b,a)$,
as it should be at the junction point. Equation (\ref{rth}) is
quartic with respect to $b^2$. For $a>256/243$ this equation
admits only one positive solution. The value of the throat's
radius as a function of the nonminimal parameter $a$ is presented
in Fig.~\ref{fig6}. When $a=256/243$, $b$ vanishes. When $a\gg 1$,
this function behaves as
$b(a)=3\left(\frac{a}{6}\right)^{1/3}-\frac{5}{12}+o(1)$.

\begin{figure}[t]
\begin{center}
\includegraphics[height=4cm]{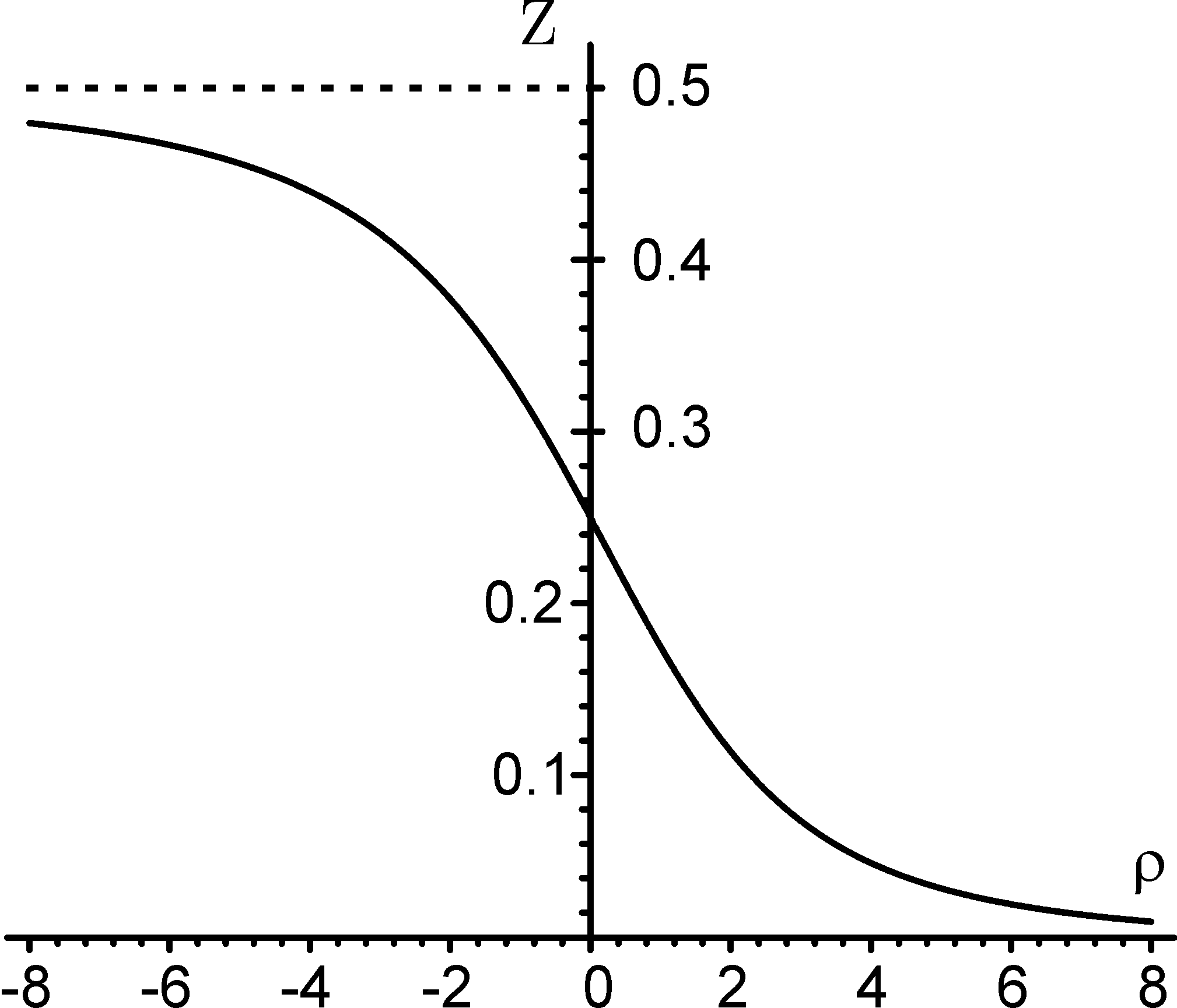}
\caption{Plot of the rescaled electric field $Z(\rho)$ for $a=4$.
The region $\rho>0$ is the Coulombian branch [$Z(+\infty)=0$], the
region of negative values of $\rho$ is the second non-Coulombian
one [$Z(-\infty)=1/2$]. The junction point $\rho=0$ is the point
of inflection on this curve.}\label{fig7}
\end{center}
\end{figure}

For illustration, it is convenient to consider the value $a=4$,
since $b(4)=2$ and $Z_{*}^{-1}(b,4)=4$ [see (\ref{Ecoul})].
Substituting $y(\rho)$ taken from (\ref{ywh}) into the cubic
equation (\ref{E}) we obtain a plot of $Z(\rho,4)$ presented in
Fig.~\ref{fig7}.
This curve is nonsymmetric: at $\rho\to +\infty$ the function
$Z(\rho)$ tends to zero (Coulombian branch), at $\rho\to -\infty$
it tends to $1/2$ (second non-Coulombian branch).

\subsection{Asymptotic behavior}

When $\rho\to +\infty$ the solution for the electric field behaves
according to the Coulombian law. In this limit the functions
$\sigma(\rho)$ and $N(\rho)$ go to one [according to
(\ref{asympCond}), (\ref{sigmaasymp}), and (\ref{Nasymp})], and we
deal with the traditional asymptotically flat Minkowski spacetime.
When $\rho\to -\infty$ the electric field $Z(\rho)$ tends to the
asymptotic value $1/\sqrt{a}\neq 0$. In this case $\sigma(\rho)$
tends to one as well, since in (\ref{s}) $\frac{dy}{d\rho}\to -1$,
while $N(\rho)$ becomes infinite
\begin{equation}
N(\rho)\sim -\frac{\rho^2}{3a}\,.
\end{equation}
Thus, for $\rho\to -\infty$ the metric of this spacetime is
asymptotically de Sitter, and the Riemann tensor in this limit may
be decomposed as
\begin{equation}
      {R^{ik}}_{mn}=-K\left(\delta^i_m\delta^k_n-\delta^i_n
\delta^k_m\right)+o(r^{-1})\,,\quad
      K=\frac{r_Q^2}{3a}=\frac{1}{6q}\,.
\end{equation}
The effective $\Lambda$-term induced by the electric field
nonminimally coupled to gravity is thus equal to $\Lambda_{{\rm
eff}} = 3 K = 1/(2q)$. In contrast to the electric strength $E$,
the electric excitation $D$ [see Eq.~(\ref{D})], being equal to
\begin{equation}
      D=\frac{Q}{Y^2}=\frac{2}{\kappa Q (\rho^2+b^2)} \,,
\end{equation}
tends to zero both for $\rho\to +\infty$ and $\rho\to -\infty$.
Similar situations, where the excitation tensor was equal to zero,
while the electric, magnetic, or other fields did not vanish, were
considered earlier in \cite{GraCos06}.

\subsection{Horizons and singularities}

\begin{figure}[h]
%\begin{center}
\includegraphics[height=3.5cm]{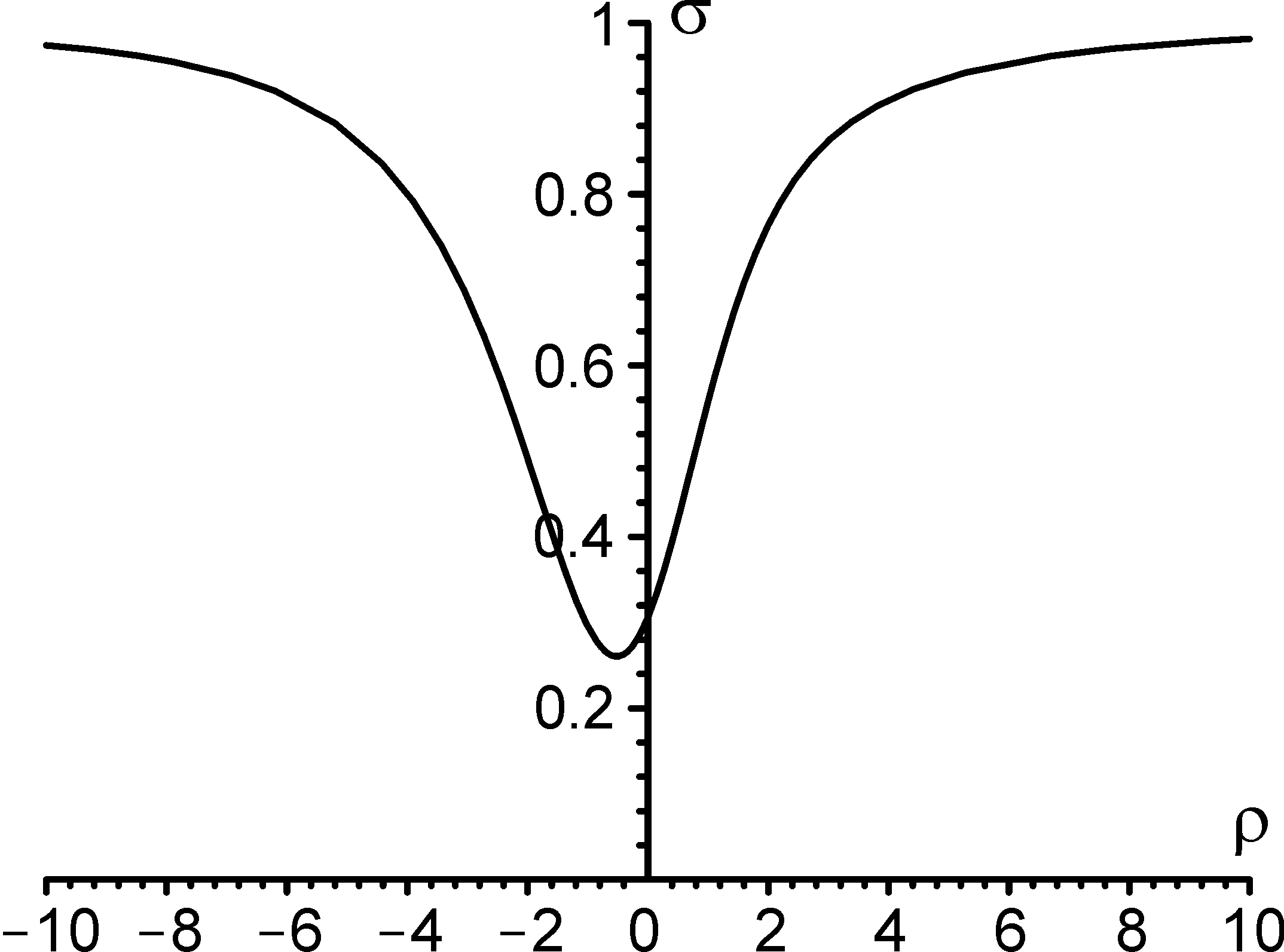}
\caption{Plot of the metric function $\sigma(\rho)$ for $a=4$.
This function is positive and tends to unity at both infinities.
The curve is nonsymmetric with respect to the vertical axis.}
%\end{center}
\label{metricsigmaw}
\end{figure}

If we intend to consider traversable wormholes only, we have to
require the spacetime metric to possess no singularities. This
means that the function $\sigma(\rho)$ should not vanish anywhere,
because, when $\sigma(\rho)=0$, we have a singularity, as
mentioned previously. Therefore, for the model considered in
\cite{BBL08} with $q_1=-q$, $q_2=q$, $q_3=0$ traversable wormholes
are absent, since in the throat $\sigma=0$ (see Eq.~(\ref{s1})).
In contrast to \cite{BBL08}, for our case ($q_1=-q$, $q_2=3q$,
$q_3=0$) we have,
\begin{equation}
   \sigma(0)=-2ab Z(0)\frac{dZ(0)}{d\rho}>0\,,
\end{equation}
because $\frac{dZ}{d\rho}<0$ at $\rho=0$ (see, e.g.,
Fig.~\ref{fig7}). However a singularity can be located at
$\rho\neq 0$.  In order to find the range for the parameter $a$
for which the function $\sigma(\rho)$ is positive everywhere, let
us consider the situation where the function $\sigma(\rho)$ given
by (\ref{s}) with $Z(\rho)$ from (\ref{E}) has a double root [in
this case the curve of $\sigma(\rho)$ touches the horizontal
axis]. Numerical calculations show that this is possible for two
values of the parameter $a$: $a=a_1\approx 1.072$ and
$a=a_2\approx 39.380$.  When $256/243< a< a_1$, the equation
$\sigma(\rho,a)=0$ has two positive roots; when $a> a_2$ the two
corresponding roots are negative. When $a_1< a< a_2$, the equation
$\sigma(\rho,a)=0$ has no real roots, i.e., $\sigma>0$ for any
value of $r$. Figure \ref{metricsigmaw} illustrates these features
for a typical value $a=4$, which is in the interval $1.072 < a <
39.380$.

\begin{figure}[h]
\halign{\hfil#\hfil&\qquad\hfil#\hfil&\qquad\hfil#\hfil\cr
\includegraphics[height=2.5cm]{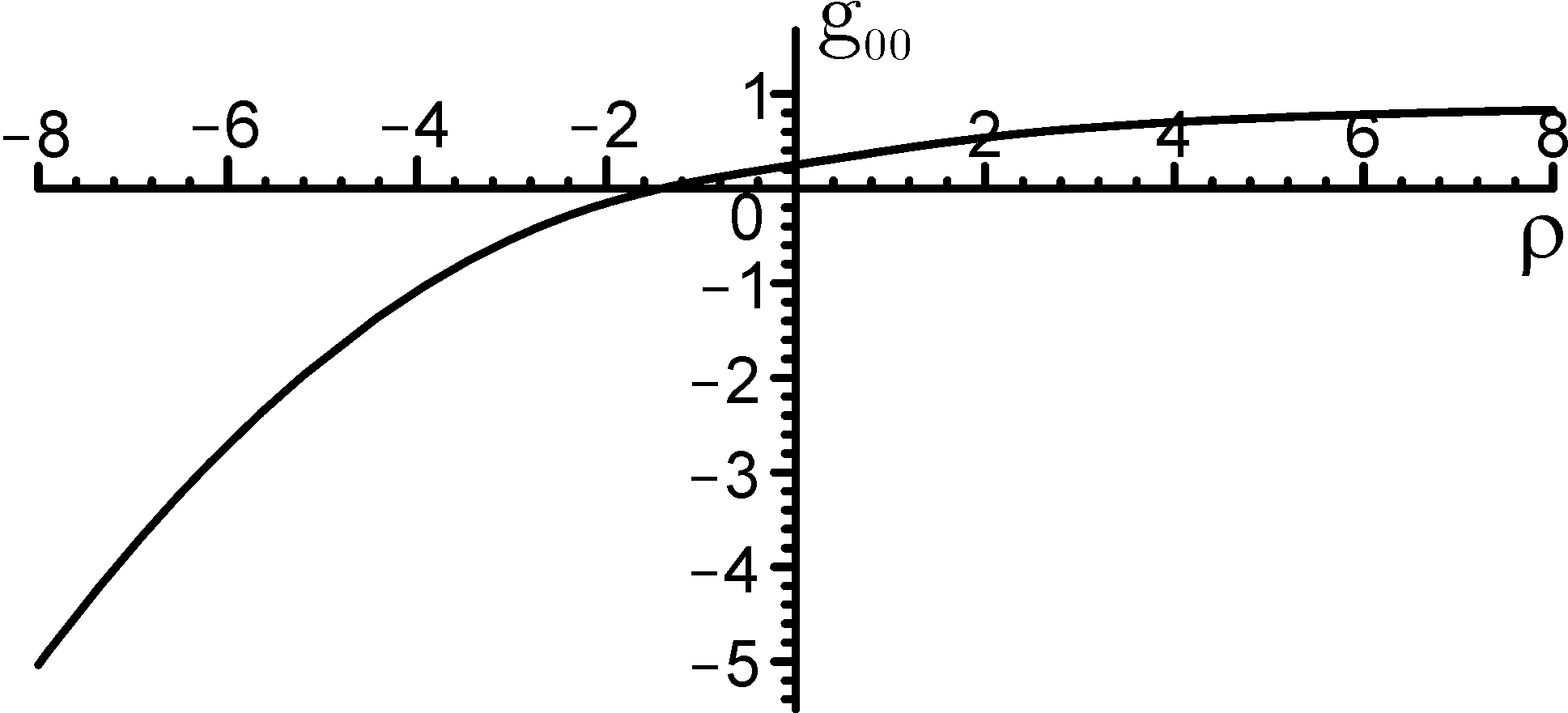}&
\includegraphics[height=2.5cm]{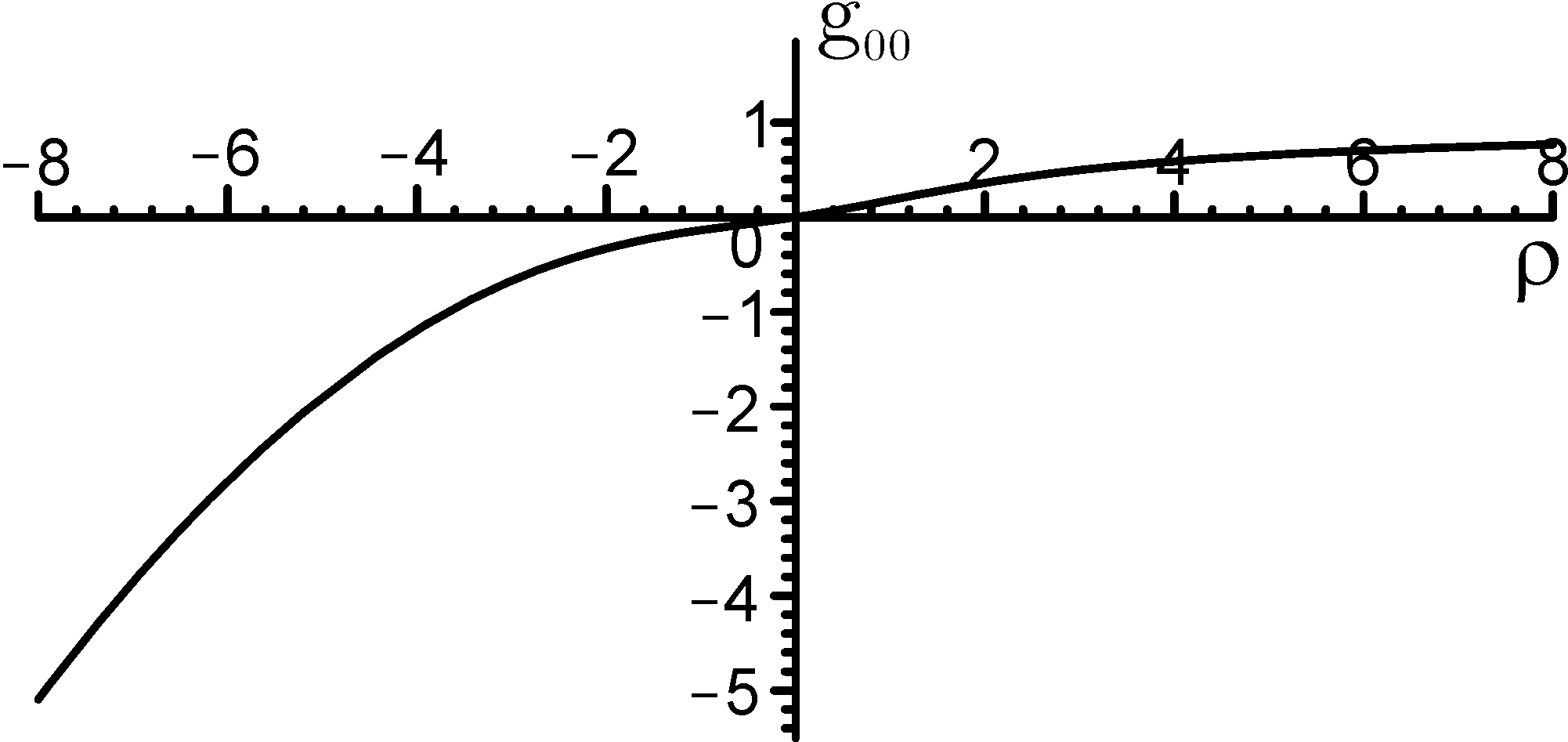}&
\includegraphics[height=2.5cm]{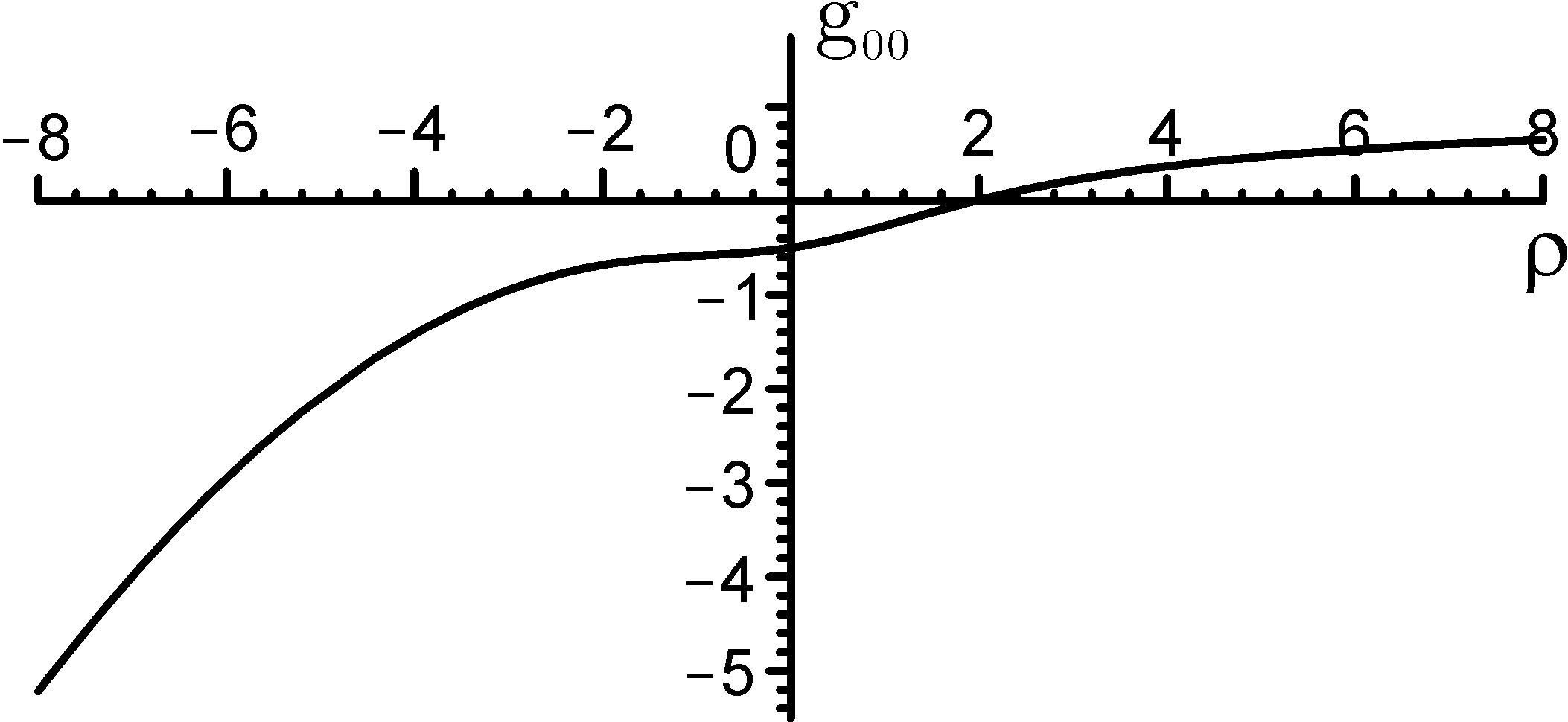}\cr
\ &\ &\ \cr
\includegraphics[height=2.5cm]{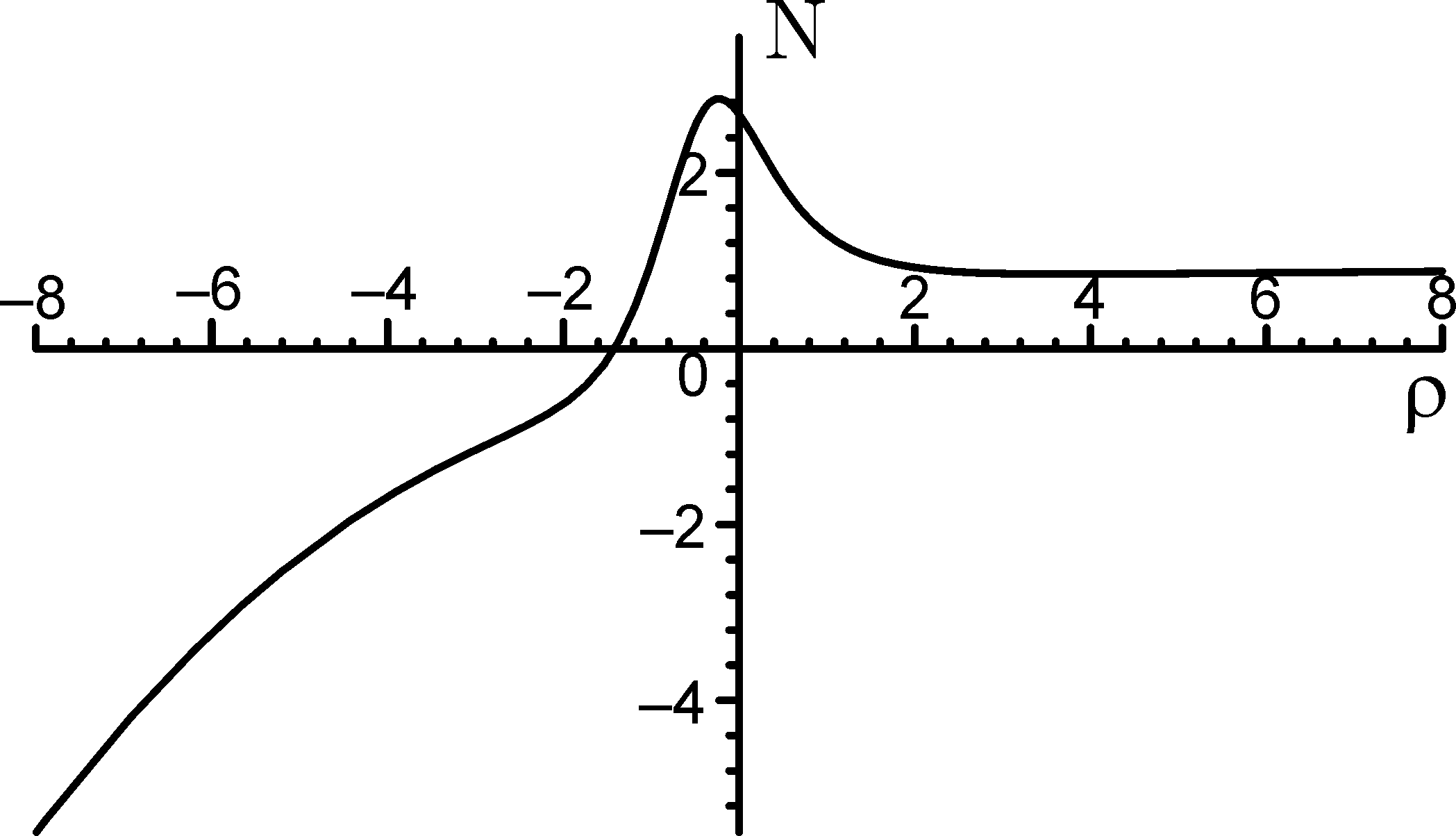}&
\includegraphics[height=2.5cm]{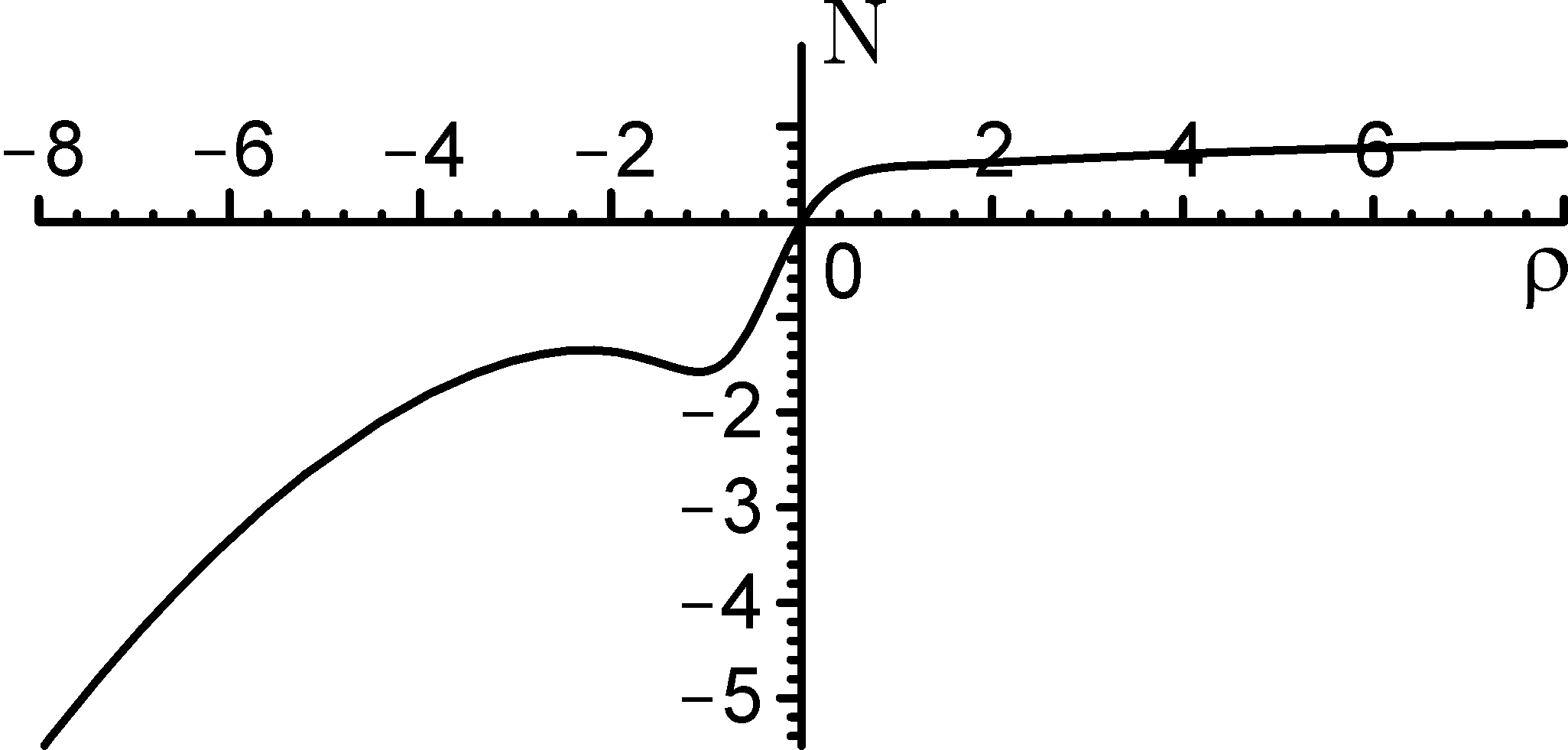}&
\includegraphics[height=3.0cm]{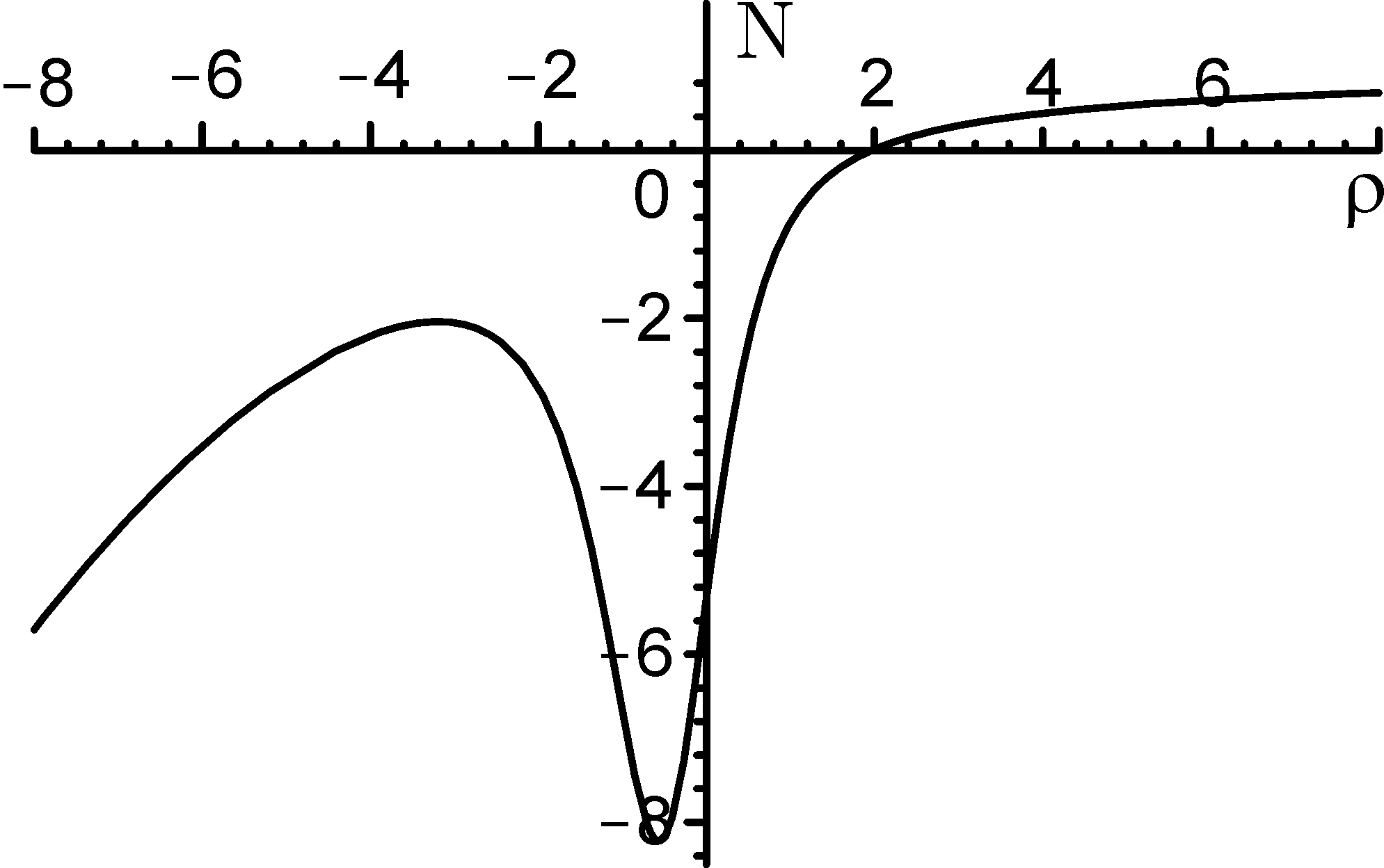}\cr
\small a) $m<\tilde{m}$ & \small b) $m=\tilde{m}=1.0081$ & \small
c) $m>\tilde{m}$ \cr} \caption{Plots of $g_{00}$ (upper row) and
$N(\rho)$ (lower row) for the value $a=4$, when there is only one
horizon. We split the interval $0<m<+\infty$ into two regions with
respect to a specific value of the rescaled mass
$\tilde{m}=1.0081$. (a) When $m<\tilde{m}$, the curves cross the
horizontal axis at negative values of $\rho$, so that the horizon
is located behind the throat. (b) When $m=\tilde{m}$, then
$g_{00}(0)=0$ and $N(0)=0$, thus, the horizon is located on the
throat. (c) When $m>\tilde{m}$, the horizon is located in front of
the throat. Plot (a) illustrates the case in which the horizon
belongs to an asymptotically de Sitter region of the combined
spacetime, it is a standard cosmological apparent horizon. In this
sense such a wormhole is traversable.} \label{metricsigmawn}
\end{figure}

Now, since the wormhole spacetime metric is an asymptotically de
Sitter one, it possesses at least one horizon (the discussion of
some aspects of wormhole physics in a cosmological context can be
found, e.g., in \cite{LambdaWH1,LambdaWH2}).  Thus, our wormhole
configuration may be considered as traversable, if, in addition to
the requirement of the absence of singularities, we suppose that
horizons of the metric are out of a traveler way through the
throat. For the metric function $N(\rho)$ it means that in the
vicinity of the throat $N(\rho)$ is positive.  To illustrate the
behavior of the functions $g_{00}(\rho)$ and $N(\rho)$ let us
consider again the model with $a=4$, see Fig.~\ref{metricsigmawn}.
When the dimensionless mass parameter $m$ is less than a critical
value $\tilde{m}$ (being equal to $1.0081$ for $a=4$), an event
horizon is located behind the throat, hence in this case the
wormhole throat is traversable.  When $m>\tilde{m}$, a horizon is
situated in front of the throat, and a traveler cannot go through,
it is a nontraversable wormhole.

\subsection{Comparisons}
It is of interest to briefly compare our electric wormhole
solutions with the magnetic nonminimal Wu-Yang wormhole solutions
studied in \cite{BaSuZa}. The magnetic nonminimal traversable
wormhole joins two asymptotically Minkowskian flat regions and the
magnetic gauge field is symmetric with respect to wormhole throat,
i.e., this field asymptotically vanishes in each region of the
combined spacetime. On the other hand, the electric nonminimal
wormhole joins an asymptotically flat region and an asymptotically
de Sitter region, for which the corresponding electric field tends
to a constant value. This introduces a new effective, nonminimally
induced, cosmological constant $\Lambda_{{\rm eff}} = 1/(2q)$.

%\newpage
\section{Conclusions}\label{V}

We have presented new exact solutions of a nonminimal
Einstein-Maxwell model with coupling parameters satisfying the
relations $q_1= -q$, $q_2=3q$, $q_3=0$, with $q$ arbitrary, which
describe spherically symmetric electrically charged objects with
and without center. The electric field of the objects is
characterized by the following features: {\it (i)} it satisfies a
cubic key equation and splits into three branches, one of them has
Coulombian asymptotics, $E(r) \to Q/r^{2}$; {\it (ii)} when the
guiding nonminimal dimensionless parameter $a=4q/\kappa Q^2$ is
positive, the electric field is described by a smooth function
finite everywhere; {\it (iii)} when $a \geq 256/243$ the
Coulombian branch of the electric field conjugates with the
non-Coulombian branch with constant asymptotics $E=1/\sqrt{\kappa
q}$, the position of the junction point depends on the parameter
$a$ and coincides with geometrical center for $a=256/243$.

There are wormhole solutions. Indeed, when the nonminimal guiding
parameter $a$ is in the interval $1.072<a<39.380$ there are no
spacetime singularities and the nonminimal field configuration
without center is a wormhole. This is an explicit example of a
nonminimal traversable electrically charged wormhole joining two
regions of spacetime, an asymptotically flat region and an
asymptotically de Sitter region.  The spacetime of this wormhole
has at least one horizon, which, depending on the value of the
rescaled asymptotic mass $m$, can be situated in front of the
throat, just on the throat and behind it. When $m$ does not exceed
some critical value (e.g., $m < 1.0081$ for $a=4$), the horizon is
located behind the wormhole throat, i.e., in the asymptotically de
Sitter region of the combined spacetime. Such a configuration is
thus a traversable wormhole supported by an electric field
nonminimally coupled to gravity.  In this manner we have presented
explicitly a nonminimal realization of Wheeler's idea about charge
without charge, and showed that, if the world is nonminimal in the
coupling of gravity to electromagnetism, then wormhole appearance
, or perhaps construction by an absurdly advanced civilization, is
possible.

Perhaps, a next natural step is to consider nonminimal
Einstein-Maxwell models with an a priori cosmological constant
$\Lambda$.  It is expect that one can also find solutions of our
nonminimal Einstein-Maxwell model.  These solution would describe
electric or magnetic wormholes joining different combinations of
asymptotically de Sitter, anti de Sitter, and Minkowski regions.
For instance, an anti de Sitter-Minkowski electric nonminimal
wormhole is expected to be free of horizons.

%\appendix

\begin{acknowledgments}
This work was partially supported by the Russian Foundation for
Basic Research, Grants Nos. 08-02-00325-a and 09-05-99015, and by
FCT - Portugal through Projects Nos. PTDC/FIS/098962/2008 and
CERN/FP/109276/2009.

\end{acknowledgments}

\end{document}